\documentclass[nofootinbib,floatfix,nolongbibliography,twocolumn]{revtex4-2}
\usepackage{amsmath}
\usepackage{amssymb}
\usepackage[version=4]{mhchem} %for molecular formula typesetting
\usepackage{siunitx}[=v2]
\usepackage{graphicx}
\usepackage{verbatim}
\usepackage{color}
\usepackage{paralist}
\usepackage{lipsum}
\usepackage{fancyhdr} % needed for header and footer
\usepackage{upgreek}
\usepackage{multirow,array,hhline,booktabs} % needed for tables

\newcommand{\ket}[1]{\left\vert{#1}\right\rangle}

\usepackage[modulo]{lineno}
\usepackage[super]{nth}

\AtBeginDocument{}
\usepackage[pdftex, breaklinks, colorlinks, citecolor=blue, urlcolor=blue]{hyperref}

\begin{document}
\author{Wonill Ha}
\author{Sieu D. Ha}
\thanks{Corresponding author. E-mail: sdha@hrl.com}
\author{Maxwell D. Choi}
\author{Yan Tang}
\author{Adele E. Schmitz}
\author{Mark P. Levendorf}
\author{Kangmu Lee}
\author{James M. Chappell}
\author{Tower S. Adams}
\author{Daniel R. Hulbert}
\author{Edwin Acuna}
\author{Ramsey S. Noah}
\author{Justine W. Matten}
\author{Michael P. Jura}
\author{Jeffrey A. Wright}
\author{Matthew T. Rakher}
\author{Matthew G. Borselli}
\affiliation{HRL Laboratories, LLC, 3011 Malibu Canyon Road, Malibu, California 90265, USA}

\begin{abstract}
Spin-based silicon quantum dots are an attractive qubit technology for quantum information processing with respect to coherence time, control, and engineering.  Here we present an exchange-only Si qubit device platform that combines the throughput of CMOS-like wafer processing with the versatility of direct-write lithography.  The technology, which we coin ``SLEDGE," features dot-shaped gates that are patterned simultaneously on one topographical plane and subsequently connected by vias to interconnect metal lines.  The process design enables non-trivial layouts as well as flexibility in gate dimensions, material selection, and additional device features such as for rf qubit control.  We show that the SLEDGE process has reduced electrostatic disorder with respect to traditional overlapping gate devices with lift-off metallization, and we present spin coherent exchange oscillations and single qubit blind randomized benchmarking data.
\end{abstract}

\title{A flexible design platform for Si/SiGe exchange-only qubits with low disorder}
\date{\today}
\maketitle

%\linenumbers

Gated silicon quantum dots are of widespread interest as physical qubits for quantum information processing due to favorable coherence times~\cite{Veldhorst2014, eng2015}, a variety of control techniques~\cite{Muhonen2015,kawakami2016,HRLAndrews2019}, and the prospect of leveraging well-established Si CMOS engineering for density, scale, and yield.  However, much of the Si qubit work to date has been demonstrated using gates defined by lift-off metallization techniques~\cite{Veldhorst2015, Zajac2016}, which suffer from poor wafer-level process control and which have not been used in Si integrated circuit foundries in several decades~\cite{homma1982liftoff}.  There have been examples of Si quantum dots fabricated in CMOS facilities with industry-standard processes~\cite{maurand2016cmos, gilbert2020twobytwo, chanrion2020twobyfour, zwerver2021intelqb}, some of which have exhibited coherent control of single electron spins. However, single-spin qubits are known to be susceptible to decoherence from global magnetic fields, among several disadvantages~\cite{russ2016}.  In this work, we present a device fabrication process termed SLEDGE (Single Layer Etch-Defined Gate Electrodes) for Si/SiGe heterostructure exchange-only spin qubits.  The SLEDGE platform uses processes designed for typical CMOS wafer fabrication, but it has inherent design flexibility for highly-customizable device layouts.  We show that the SLEDGE process has reduced electrostatic disorder with respect to lift-off technologies, and we show representative coherent exchange oscillations and single qubit blind randomized benchmarking data to extract quantum gate performance~\cite{HRLAndrews2019}.

\begin{figure}[th]
\includegraphics[width=.9\linewidth]{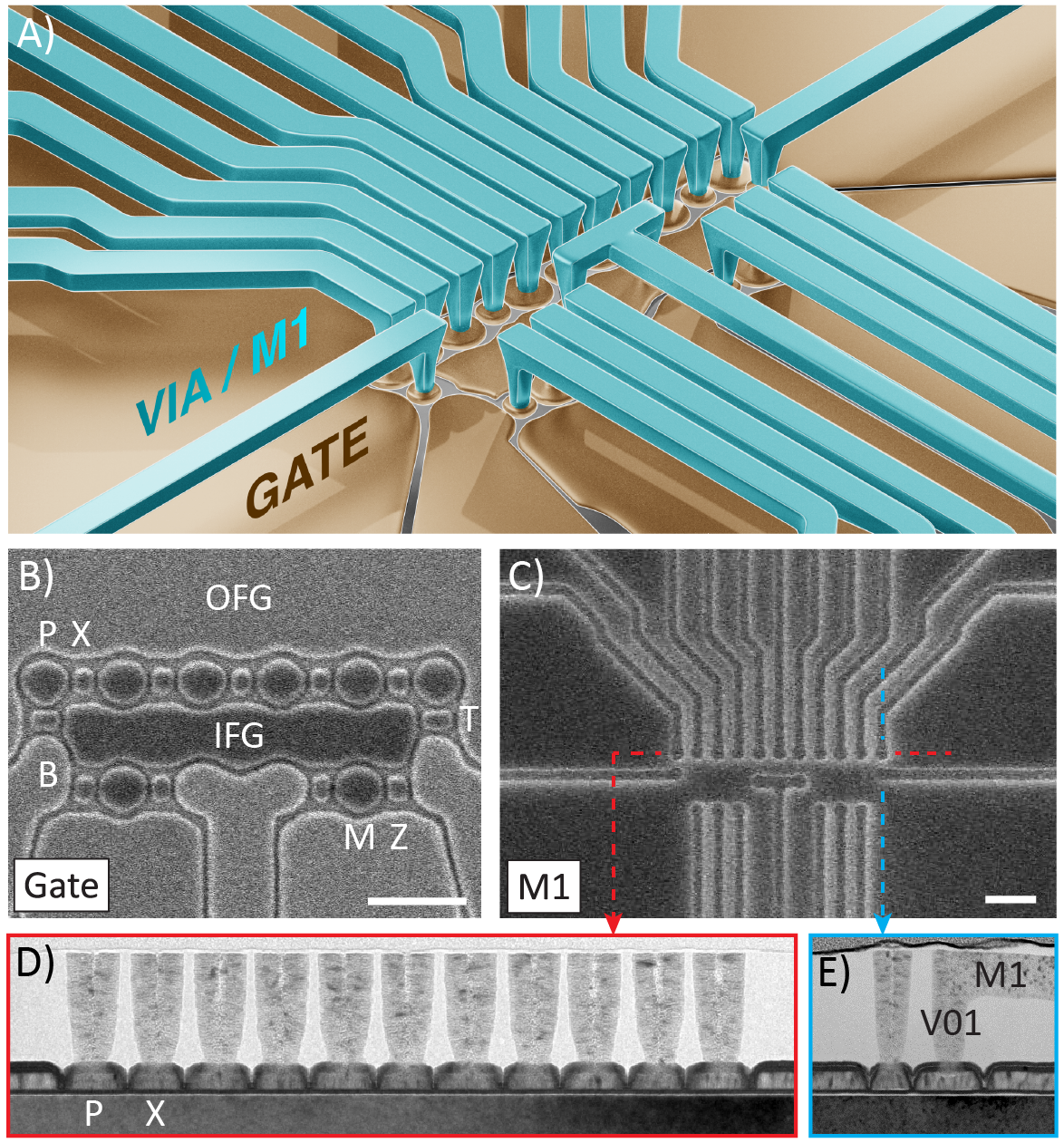}
\caption{
A) Illustrated render of SLEDGE device. Gate level in tan and BEOL levels in aqua.  B) Top-down SEM image after gate etch with gaps evident between gates. Labels are as described in text. C) Top-down SEM image after M1 CMP of dual damascene BEOL. Spans connect each via to bond-level routing (not shown). Scale bars in (B) and (C) are \SI{200}{\nm}. D) Cross-section TEM image through P-gate row, as illustrated by red dashed line in (C), showing BEOL vias contacting uniplanar gates.  E) Cross-section TEM image through right-most P-T gates, as illustrated by blue dashed line in (C), showing M1 trench in ILD connecting to P-gate via.
}
\label{figdev}
\end{figure}

\begin{figure}
\includegraphics[width=1\linewidth]{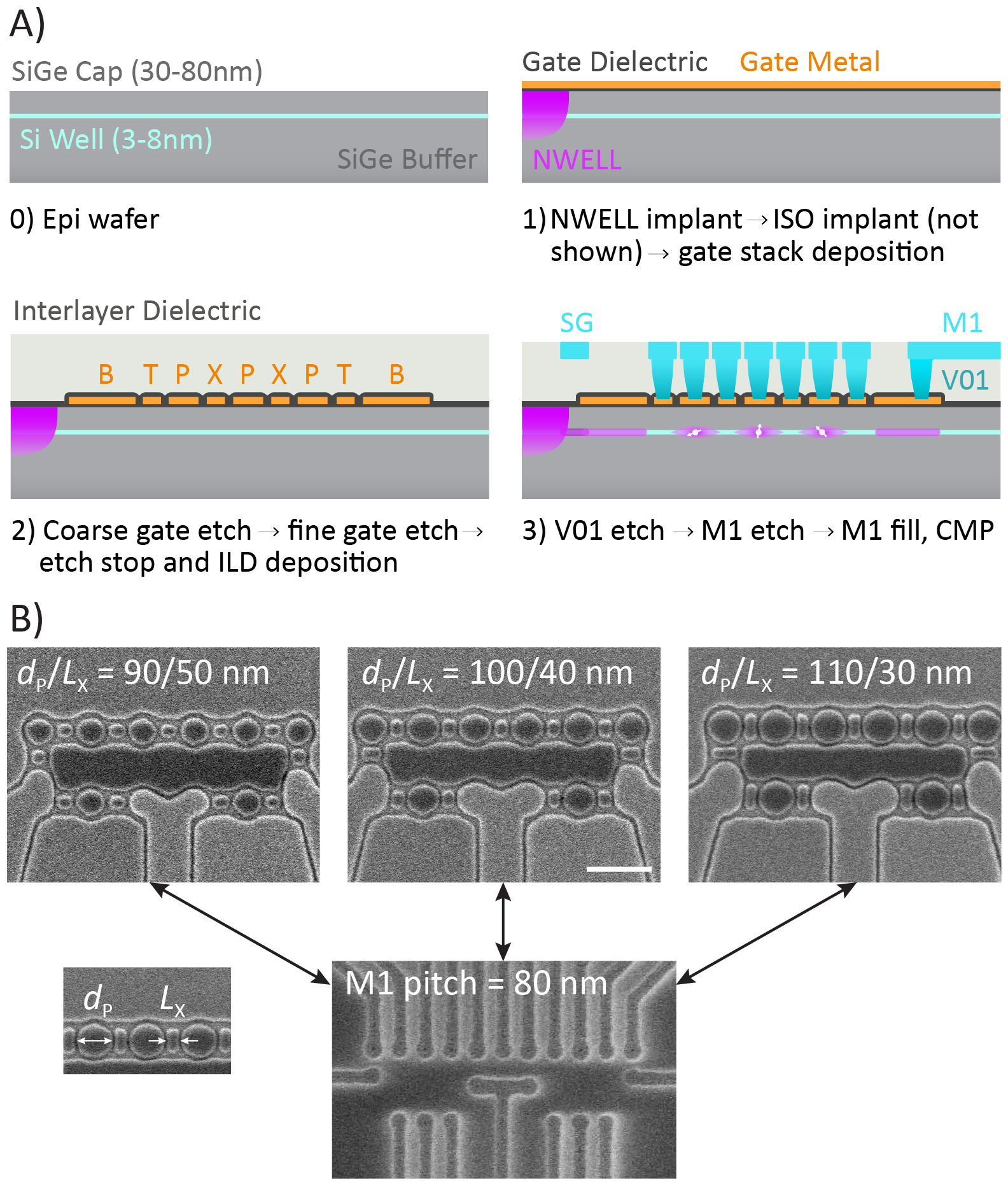}
\centering
\caption{A) Depiction of SLEDGE process flow, not to scale. Cross-section is an illustrative piecewise cut from an Ohmic, through bath gates, up T-gates, and across a triple dot P-gate row. Steps are as described in text. ``SG" in lower right panel is supply gate patterned at M1.  B) Top-down SEM images of three different devices patterned on the same wafer, each with varying P- and X-gate dimensions.  The P-gate diameter ($d_\textup{P}$) and X-gate length ($L_\textup{X}$) parameters are defined in the lower left image with \SI{10}{\nm} gaps between gates.  As each device has the same P-P pitch and M-P distance, all use the same M1-level pattern as shown in the lower SEM image.  All images are at the same scale, with \SI{200}{\nm} scale bar in center top image. 
}
\label{figflow}
\end{figure}

The SLEDGE technology is designed for heterostructure quantum dots, wherein semiconductor epitaxial boundaries and gate-defined electric fields confine and, in the latter case, manipulate individual electrons.  A SLEDGE device, as shown in Fig.~\ref{figdev}(a), consists of 1) a gate layer, in which all gates are patterned simultaneously on the same plane (i.e. uniplanar), and 2) back-end-of-line (BEOL) interconnect layers, in which vertical vias directly contact active gates and then spans (``M1") connect vias to macroscopic routing \SIrange{10}{50}{\um} away (not shown).  The gate-level design (Fig.~\ref{figdev}(b)) is similar to recent work on Si quantum dot devices~\cite{Zajac2016, HRLJones2019}.  Plunger (P) and exchange (X) gates for accumulating and exchange-coupling electron spins are arranged collinearly and are offset from measure dot (M) gates used for charge sensing and readout of spin-to-charge conversion.  Electrons are supplied by bath (B) gates via tunnel gates (T and Z) to P- and M-gates, respectively.  Electron baths are supplied from source/drain Ohmics via supply gates (SG, see Fig.~\ref{figflow}(a) lower right), which control Ohmic-bath contact resistance independent of B-gate voltage~\cite{vanbeveren2010supplygate}.  There are two field gates used for depleting carriers around the active gates: the inner field gate (IFG) for the region between P and M gates, and the outer field gate (OFG) for the periphery.  SLEDGE devices are nominally designed for triple-dot exchange-only qubit operation~\cite{divincenzo2000}, but could be used for a variety of encodings.

The uniplanar gate arrangement is a key feature of SLEDGE and is unlike conventional quantum dot qubit devices that have been demonstrated to date, in which exchange gates (sometimes referred to as barrier gates) are on a separate dielectric plane and may overlap laterally with plunger gates~\cite{lawrie2020quantum}.  From a qubit operation perspective, exchange gates on the same topographical plane as plunger gates experience less electric field screening than those on a separate plane and therefore require smaller voltage throws to modulate exchange energy for qubit rotations (see supplement). The absence of an additional dielectric layer between exchange gates and quantum well should also reduce charge noise~\cite{connors2019noise}.  In addition to the uniplanar gate design, the quantum dot gates are notably dot-shaped ($\sim$zero-dimensional), as opposed to conventional extended line-shaped ($\sim$one-dimensional) gates.  The major drawback of using one-dimensional gate lines to define a zero-dimensional quantum dot is that device designs are essentially topologically limited to ring-like (i.e. two-row) geometries, whereas SLEDGE designs can be extended to highly-customizable gate arrangements with additional BEOL levels as needed (see supplement).

The process flow to fabricate SLEDGE devices on Si/SiGe quantum well heterostructures is outlined in Fig.~\ref{figflow}(a).  The heterostructure (``epi wafer'') consists of a tensile-strained Si quantum well epitaxially grown on a strain-relaxed $\textup{Si}_{1-x}\textup{Ge}_x$ ($x=0.25-0.35$) buffer, followed by a SiGe capping layer of the same stoichiometry as the buffer. The beginning of the process flow includes optical lithography steps to pattern degenerately doped phosphorus-implanted Ohmics (NWELL) and argon-implanted electrically inactive regions (ISO).  After defining implant regions, the gate dielectric (bilayer \ce{Al2O3}/\ce{HfO2}) and gate metal (TiN) stack are blanket deposited, and gates are patterned in two steps.  First, an optically-defined coarse etch removes gate metal from a majority of the wafer, leaving a block of gate metal for each device from which all active gates will be subtractively patterned.  Second, positive tone e-beam lithography is used to write gaps between gates, and a F-based dry-etch of gaps defines gates, as seen in the top-down SEM image of Fig.~\ref{figdev}(b).  The process flow then enters the BEOL phase, wherein gates are contacted by vias (V01, connects M0 (gate) to M1) through an interlayer dielectric material (ILD, \ce{SiO2}), and spans (M1) connect vias to macroscopic routing at the bond pad level. A top-down SEM image after M1 patterning is shown in Fig.~\ref{figdev}(c).

The SLEDGE gate process is compatible with a variety of BEOL integration schemes due to its topographically flat design.  Fig.~\ref{figdev}(d-e) shows cross-sectional TEM images along horizontal and vertical cuts from a TiN dual damascene process.  (We have also demonstrated a subtractive BEOL process flow, as discussed in the supplement).  Here, vias are patterned by e-beam lithography and etched into the ILD, stopping selectively on a \ce{HfO2} etch stop dielectric layer deposited on top of the gates prior to ILD deposition.  M1 spans are then patterned by e-beam lithography and trenches are etched into the ILD.  Following e-beam resist strip, a blanket etch completes the via etch down to gates through the etch stop layer.  We fill V01 and M1 with an atomic layer deposition (ALD) TiN process, followed by chemical-mechanical polishing (CMP) to remove excess TiN and isolate M1 spans (Fig.~\ref{figdev}(c)).  M1 is subsequently contacted by an optically-defined metallization step to create wide (\SIrange{1}{3}{\um}) routing lines and bond pads for wirebonding to a chip carrier.  

Using direct-write e-beam lithography with separate gate and BEOL contact layers allows for within-wafer flexibility in device design. In one reticle (\SI{\sim10 x 10}{\mm}), of which many are patterned across each wafer, we can fabricate devices with varying parameters to explore a broad design space including, but not limited to, gate pitch, P-M separation, relative P/X ratio, and M1 width.  The gate, V01, and M1-level designs can be easily and independently modified in layout.  Using appropriate proximity effect correction~\cite{hauptmann2009pec}, new designs are fractured and patterned in e-beam lithography without needing substantial development, if any, to optimize exposure or develop conditions.  As an example, we show in Fig.~\ref{figflow}(b) SEM images of three different devices patterned in one reticle with varying P/X dimensions but fixed pitch and fixed V01/M1-level design.  The P-gate diameter ($d_\textup{P}$) and X-gate length ($L_\textup{X}$) are varied from \SI{90/50}{\nm} to \SI{110/30}{\nm}, but the BEOL dimensions are unchanged between devices.  This allows for examining relative P- and X-gate capacitances without confounding effects from differences in contact layers.  The device design flexibility is additionally independent of the substrate heterostructure design, wherein Si well and SiGe cap can be modified in the epitaxial growth stage prior to fabrication. The ability to explore a broad parameter space is crucial for converging on the optimal device design for spin-based quantum information processing.

Flexibility of the SLEDGE process extends beyond gate-level device design.  BEOL materials can be chosen separately from gate metal provided a sufficient Ohmic contact can be engineered.  For example, a subtractive BEOL process can be used if sputtered metals (e.g. Nb or Pt) are desired, while dual damascene is conducive to materials deposited by chemical vapor deposition or ALD (e.g. W or TiN).  Moreover, the BEOL span fanout can be readily designed to allow space for additional active features in arbitrary locations around the device, such as a micromagnet or a superconducting resonator~\cite{kawakami2016, mi2017cqed}, again independent of the gate-level design (see supplement).

The requirements for e-beam overlay between V01-gate and M1-V01 layers are approximately \SIrange{5}{15}{\nm}, set by gate pitch and minimum gate dimension.  While non-trivial, we readily achieve the alignment requirements within-wafer and across lots with a non-customized, commercially available Raith EBPG5200 e-beam writer.  In the supplemental materials, we show that the mean misalignment magnitude across four representative lots (with four wafers per lot) is \SI{\lesssim5}{\nm} for both V01-gate and M1-V01 levels.   We use the tool with four \SI[number-unit-product=\text{-}]{3}{in} wafers (i.e. one lot) mounted onto a \SI{200}{\mm} wafer holder, and therefore the overlay we achieve represents a \SI{200}{\mm} wafer capability for e-beam lithography.  We show also that the mean P-gate critical dimension (CD) in lithography varies $<7\%$ between the same four lots.  Overall, this demonstrates that e-beam lithography is well capable of wafer-level Si qubit fabrication. 

In addition to device design flexibility, the SLEDGE process flow enables aggressive SiGe surface cleaning prior to gate dielectric deposition.  With overlapping gate technologies \cite{Zajac2016,eenink2019mos}, the screening gate (also known as confinement gate) and underlying gate dielectric, which are commonly Al-based, preclude most wet surface cleans prior to plunger gate dielectric deposition.  Similar restrictions exist for exchange gate dielectric deposition. This can lead to substantial interface defect densities and associated electrostatic non-uniformity.  Pd gates are more etch-resistant than Al~\cite{brauns2018pd}, but still may be roughened or delaminated in common HCl- or piranha-based (\ce{H2SO4}:\ce{H2O2}:\ce{H2O}) cleaning chemistries.  In the SLEDGE process with simultaneously patterned uniplanar gates, there is only one dielectric layer for all gates, and the semiconductor surface is free of metal and dielectric materials prior to dielectric deposition. Therefore, wet etchants for trace metal, oxide, and carbon contamination, such as SC1 (\ce{NH4OH}:\ce{H2O2}:\ce{H2O}), SC2 (HCl:\ce{H2O2}:\ce{H2O}), HF, and piranha can be used given ratios are chosen to have reasonable selectivity to SiGe. 

The improvement in electrostatic uniformity with appropriate surface cleaning can be demonstrated from characteristics of gated Hall bars (HB) fabricated on-wafer alongside quantum dot devices.  On our device wafers, gates for HBs (\SI{\sim400 x 20}{\um}) are fabricated as process control monitors at each planar metallization step (e.g. gate, M1, bond), and Hall parameters are measured to characterize the epitaxial material and device dielectric layers.  Such HBs serve to quickly quantify intra- and inter-wafer gate/semiconductor uniformity due to relative ease of fabrication and measurement interpretation.  We use low carrier density HB properties to quantify potential disorder~\cite{efros1993}, which influences charge manipulation at the dot level.  In two-dimensional electron systems, potential disorder on relevant length scales can be characterized by the density at which there is a crossover from metallic to insulating behavior.  The crossover density has various potential mechanisms, for example strong localization or that of classical percolation theory, which each have several measurement methods~\cite{dassarma2014}.  In our Hall measurements, we characterize disorder instead using the sheet density at which the Hall mobility extrapolates to 0 in a linear mobility-density plot, which we denote as $n_\textup{min}$.  To extract $n_\textup{min}$, we perform a line fit using only data in the low-density regime ($n$\SI{\leq4d10}{\per\square\cm}).  While not necessarily physical (negative values are possible), the advantage of $n_\textup{min}$ is that extrapolation is relatively straightforward, in contrast to the percolation density or metal-insulator transition density.  The latter may involve temperature sweeps and/or fitting many data points to some functional form, which can be difficult for imperfect gate or epitaxial structures.  In addition, we have found in our HBs that $n_\textup{min}$ typically scales similarly as the percolation and metal-insulator transition densities.

\begin{figure}[t]
\includegraphics[width=1\linewidth]{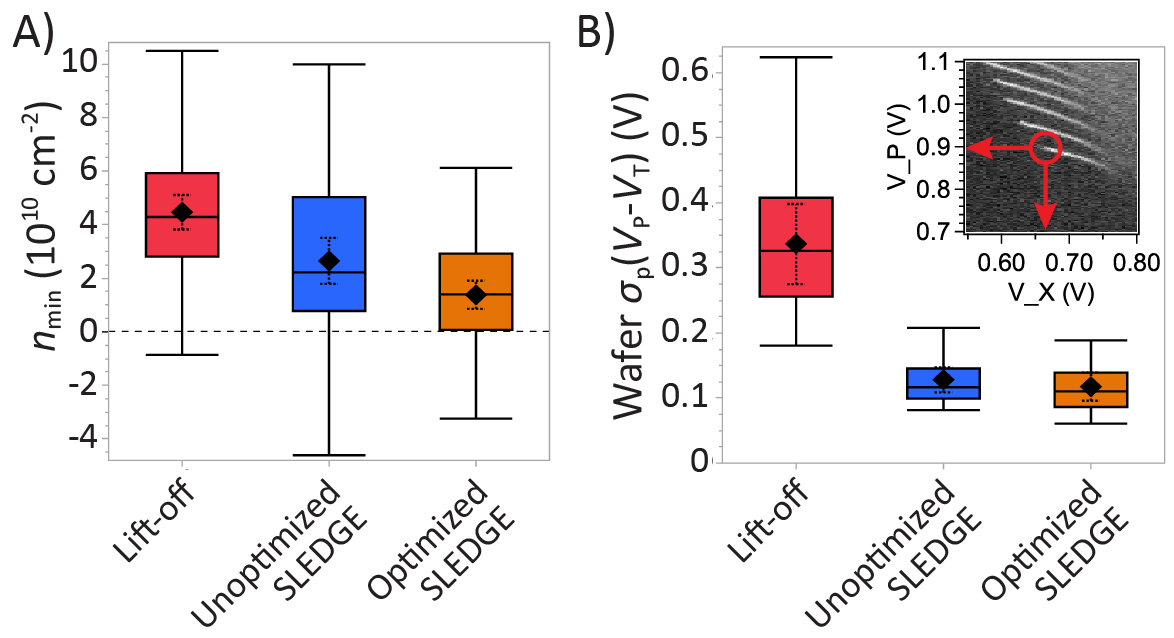}
\caption{A) Gated Hall bar measurements of $n_\textup{min}$ for various gate types.  Quartile box plots, means (diamonds), and 99\% confidence intervals (dashed whiskers) are shown.  Mean $n_\textup{min}$ exhibits a statistically significant decrease from lift-off gates to unoptimized SLEDGE gates to optimized SLEDGE gates. B) Per-wafer pooled standard deviations of P- and X-/T-gate voltage differences at first electron loading line in single dot charge stability diagram (inset), measured at \SI{1.6}{\K}.  Quartile box plots, means, and 99\% confidence intervals are shown.  There is a statistically significant decrease in mean from lift-off gates to SLEDGE gates, unoptimized or optimized.  
}
\label{figDisord}
\end{figure}

In Fig.~\ref{figDisord}(a), we plot $n_\textup{min}$ distributions for lift-off plunger gates (from our overlapping gate technology prior to SLEDGE), SLEDGE gates with unoptimized surface cleaning, and SLEDGE gates with optimized surface cleaning.  From lift-off gates to unoptimized SLEDGE gates, mean $n_\textup{min}$ is reduced from \SIrange{4.5d10}{2.6d10}{\per\square\cm}.  There is a further reduction in mean $n_\textup{min}$ with optimized SLEDGE gates to \SI{1.4d10}{\per\square\cm}.  The means are statistically different with $>99\%$ confidence.  The data implies a reduction in disorder by switching from overlapping or multi-plane gates, which have multiple dielectric depositions before gate patterning and limited pre-gate clean options, to uniplanar gates, which allow for more aggressive pre-deposition wet cleans.  

We can further quantify potential disorder at the device level by analyzing single dot charge stability diagrams across wafers.  One metric is to use the differential voltage between P- and neighboring T- or X-gates ($V_\textup{P}-V_\textup{T}$) at the first electron loading line, as shown in the inset of Fig.~\ref{figDisord}(b).  We have found that the differential voltage scales exponentially with SiGe cap thickness as expected from solving Laplace's equation for a pinned surface gate potential model~\cite{davies1995}.  This indicates that 1) across devices, the quantum well is tuned to approximately the same potential at the one-electron loading line, and 2) variations in the voltage difference to load one electron are due to potential disorder.  For each wafer, we find the standard deviation (pooled by device) of all ($V_\textup{P}-V_\textup{T}$) differential biases from the wafer, which we use as a metric for disorder.  In Fig.~\ref{figDisord}(b), we plot distributions of the pooled standard deviations ($\sigma_\textup{p}$) for lift-off gates, SLEDGE gates with unoptimized pre-clean, and SLEDGE gates with optimized pre-clean.  Only data from wafers with the same \SI{60}{\nm} SiGe cap thickness are plotted. As with $n_\textup{min}$, there is a statistically significant reduction ($>99\%$ confidence) in mean $\sigma_\textup{p}$ from lift-off gates to unoptimized SLEDGE gates of \SIrange{0.34}{0.13}{\V}.  The mean for optimized SLEDGE gates is \SI{0.12}{\V}, but it is not statistically different from that of unoptimized SLEDGE gates at the same confidence level, perhaps because of residual disorder in the gate stack itself.  Regardless, the HB and charge stability diagram data both show a clear reduction in wafer-level electrostatic disorder for SLEDGE devices as compared to lift-off devices.  Moreover, the SLEDGE $\sigma_\textup{p}$ means compare favorably to corresponding addition voltages (\SI{\sim40}{\mV}, see \ref{figDisord}(b) inset). 

\begin{figure*}
\includegraphics[width=1\textwidth]{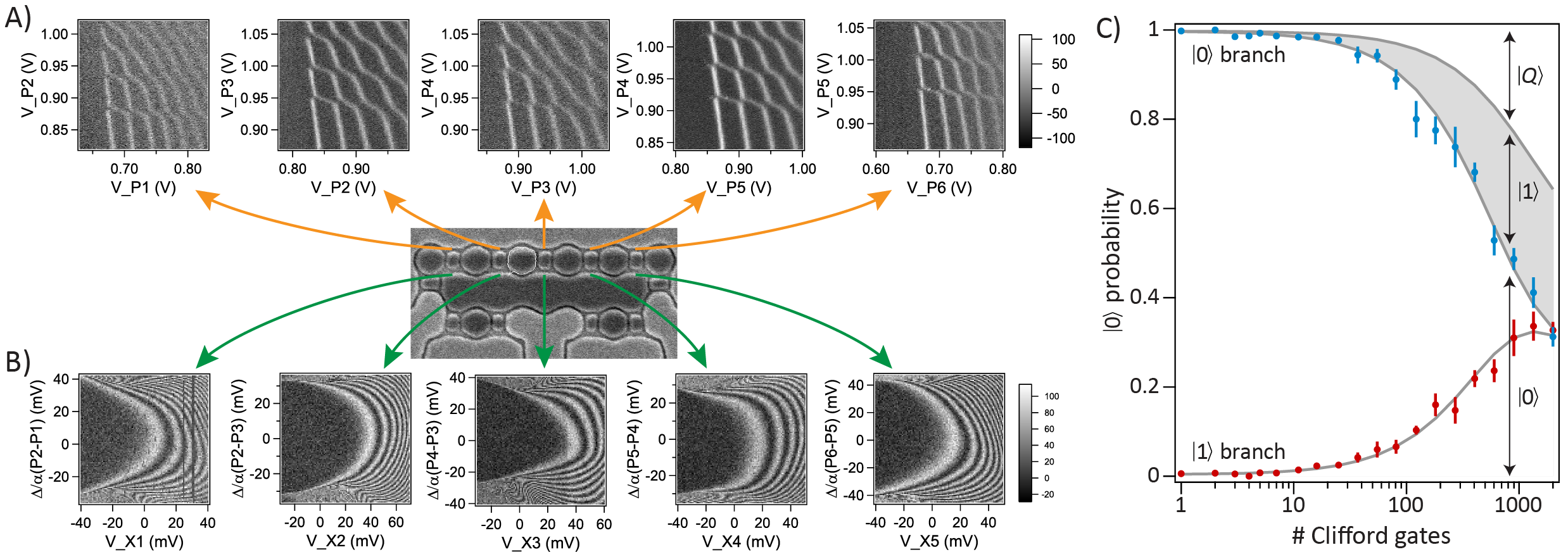}
\centering
\caption{A) Charge stability diagrams measured at \SI{1.6}{\K} from each double dot pair of a representative 6-dot SLEDGE device.  B) Exchange oscillation fringes measured in a DF for each double dot pair as a function of corresponding P-P detuning and X-gate voltage (``fingerprint" plots). Fingerprints were measured from same device as in (A)  C) Single qubit blind randomized benchmarking from a SLEDGE device. The $\ket{0}$ branch (blue points) and $\ket{1}$ branch (red points) are used to extract per-Clifford error and leakage rates. The arrows show estimated populations in states $\ket{0}$, $\ket{1}$ (shaded region), and leaked states $\ket{Q}$.
}
\label{figcryo}
\end{figure*}

Beyond the advantages in device design flexibility and disorder with the SLEDGE process, the technology also readily produces qubit devices capable of spin coherent operations.  A representative set of charge stability diagrams from a 6-dot SLEDGE device recorded at \SI{1.6}{\K} is shown in Fig.~\ref{figcryo}(a), with one diagram for each adjacent plunger gate pair. Each diagram exhibits the canonical double dot honeycomb pattern with cells merging at higher electron occupancy due to tunnel barrier lowering \cite{Borselli2015}.  Loading voltages for the most relevant (0,1), (1,0), and (1,1) charge configurations are all \SI{\le1}{\V}, and, with the exception of the outer dots, which experience enhanced cross-capacitance from the OFG, all loading voltages are similar to each other.  This is a critical feature for proposed multiplexed cryogenic control~\cite{xue2020horseridge}.

Representative plots from each X-gate showing exchange ($J$) fringes as a function of neighboring P-P detuning ($\Delta$, ordinate) and exchange gate voltage ($V_\textup{X}$, abscissa) are shown in Fig.~\ref{figcryo}(b).  In these so-called ``fingerprint" plots~\cite{reed2016}, which were measured at dilution refrigerator (DF) temperatures from the same device as in Fig.~\ref{figcryo}(a), the exchange frequency between inter-dot electrons increases with both tunnel coupling (controlled by $V_\textup{X}$) and $\Delta$.  Fingerprint plots are used to determine the symmetric axis (vector in $\Delta{\text -}V_\textup{X}$ voltage space where $\frac{dJ}{d\Delta}\sim0$) for reduced sensitivity to charge noise, and they indicate that a given double dot pair exhibits spin coherent exchange oscillations. Furthermore, single qubit blind randomized benchmarking (RB) data from an exchange-only encoded 3-dot SLEDGE device is shown in Fig.~\ref{figcryo}(c).  In the blind RB sequence, the qubit is initialized in the spin singlet ($\ket{0}$), a sequence of $N$ gates randomly selected from the Clifford group is applied, and a recovery Clifford returns the qubit ideally to the spin singlet or spin triplet ($\ket{1}$)~\cite{HRLAndrews2019}.  The sum and difference of the $\ket{0}$ and $\ket{1}$ return probabilities as a function of $N$ can be fit to exponential forms to extract per-Clifford error and leakage rates.  For this particular device, we find per-Clifford error of $0.12\pm0.01\%$ and leakage of $0.035\pm0.011\%$.

In conclusion, we have demonstrated a flexible fabrication process for Si/SiGe exchange-only qubits wherein dot-shaped gates and routing are defined on separate planes and are contacted with interconnect vias.  The separation, in combination with electron-beam lithography, enables high customizability in materials and device designs, particularly toward non-conventional layouts.  SLEDGE qubit deviecs have uniplanar gates with low electrostatic disorder compared to overlapping gate devices fabricated with lift-off metallization processes.  Future work includes improved rf engineering of M1 and optical layer routing, as well as integration of micromagnets and superconducting resonators.

\begin{acknowledgments}
The authors thank Jacob Z. Blumoff, Christopher D. Bohn, John B. Carpenter, Catherine M. Erickson, A. Mitch Jones, Maggy L. Lau, Se\'{a}n M. Meenehan, and Matthew D. Reed for valuable discussions and contributions.
\end{acknowledgments}

\bibliography{qdotref}

\begin{thebibliography}{28}%
\makeatletter
\providecommand \@ifxundefined [1]{%
 \@ifx{#1\undefined}
}%
\providecommand \@ifnum [1]{%
 \ifnum #1\expandafter \@firstoftwo
 \else \expandafter \@secondoftwo
 \fi
}%
\providecommand \@ifx [1]{%
 \ifx #1\expandafter \@firstoftwo
 \else \expandafter \@secondoftwo
 \fi
}%
\providecommand \natexlab [1]{#1}%
\providecommand \enquote  [1]{``#1''}%
\providecommand \bibnamefont  [1]{#1}%
\providecommand \bibfnamefont [1]{#1}%
\providecommand \citenamefont [1]{#1}%
\providecommand \href@noop [0]{\@secondoftwo}%
\providecommand \href [0]{\begingroup \@sanitize@url \@href}%
\providecommand \@href[1]{\@@startlink{#1}\@@href}%
\providecommand \@@href[1]{\endgroup#1\@@endlink}%
\providecommand \@sanitize@url [0]{\catcode `\\12\catcode `\$12\catcode
  `\&12\catcode `\#12\catcode `\^12\catcode `\_12\catcode `\%12\relax}%
\providecommand \@@startlink[1]{}%
\providecommand \@@endlink[0]{}%
\providecommand \url  [0]{\begingroup\@sanitize@url \@url }%
\providecommand \@url [1]{\endgroup\@href {#1}{\urlprefix }}%
\providecommand \urlprefix  [0]{URL }%
\providecommand \Eprint [0]{\href }%
\providecommand \doibase [0]{https://doi.org/}%
\providecommand \selectlanguage [0]{\@gobble}%
\providecommand \bibinfo  [0]{\@secondoftwo}%
\providecommand \bibfield  [0]{\@secondoftwo}%
\providecommand \translation [1]{[#1]}%
\providecommand \BibitemOpen [0]{}%
\providecommand \bibitemStop [0]{}%
\providecommand \bibitemNoStop [0]{.\EOS\space}%
\providecommand \EOS [0]{\spacefactor3000\relax}%
\providecommand \BibitemShut  [1]{\csname bibitem#1\endcsname}%
\let\auto@bib@innerbib\@empty
%</preamble>
\bibitem [{\citenamefont {Veldhorst}\ \emph {et~al.}(2014)\citenamefont
  {Veldhorst}, \citenamefont {Hwang}, \citenamefont {Yang}, \citenamefont
  {Leenstra}, \citenamefont {de~Ronde}, \citenamefont {Dehollain},
  \citenamefont {Muhonen}, \citenamefont {Hudson}, \citenamefont {Itoh},
  \citenamefont {Morello},\ and\ \citenamefont {Dzurak}}]{Veldhorst2014}%
  \BibitemOpen
  \bibfield  {author} {\bibinfo {author} {\bibfnamefont {M.}~\bibnamefont
  {Veldhorst}}, \bibinfo {author} {\bibfnamefont {J.~C.~C.}\ \bibnamefont
  {Hwang}}, \bibinfo {author} {\bibfnamefont {C.~H.}\ \bibnamefont {Yang}},
  \bibinfo {author} {\bibfnamefont {A.~W.}\ \bibnamefont {Leenstra}}, \bibinfo
  {author} {\bibfnamefont {B.}~\bibnamefont {de~Ronde}}, \bibinfo {author}
  {\bibfnamefont {J.~P.}\ \bibnamefont {Dehollain}}, \bibinfo {author}
  {\bibfnamefont {J.~T.}\ \bibnamefont {Muhonen}}, \bibinfo {author}
  {\bibfnamefont {F.~E.}\ \bibnamefont {Hudson}}, \bibinfo {author}
  {\bibfnamefont {K.~M.}\ \bibnamefont {Itoh}}, \bibinfo {author}
  {\bibfnamefont {A.}~\bibnamefont {Morello}},\ and\ \bibinfo {author}
  {\bibfnamefont {A.~S.}\ \bibnamefont {Dzurak}},\ }\href
  {http://dx.doi.org/10.1038/nnano.2014.216} {\bibfield  {journal} {\bibinfo
  {journal} {Nature Nanotechnology}\ }\textbf {\bibinfo {volume} {9}},\
  \bibinfo {pages} {981 EP } (\bibinfo {year} {2014})}\BibitemShut {NoStop}%
\bibitem [{\citenamefont {Eng}\ \emph {et~al.}(2015)\citenamefont {Eng},
  \citenamefont {Ladd}, \citenamefont {Smith}, \citenamefont {Borselli},
  \citenamefont {Kiselev}, \citenamefont {Fong}, \citenamefont {Holabird},
  \citenamefont {Hazard}, \citenamefont {Huang}, \citenamefont {Deelman},
  \citenamefont {Milosavljevic}, \citenamefont {Schmitz}, \citenamefont {Ross},
  \citenamefont {Gyure},\ and\ \citenamefont {Hunter}}]{eng2015}%
  \BibitemOpen
  \bibfield  {author} {\bibinfo {author} {\bibfnamefont {K.}~\bibnamefont
  {Eng}}, \bibinfo {author} {\bibfnamefont {T.~D.}\ \bibnamefont {Ladd}},
  \bibinfo {author} {\bibfnamefont {A.}~\bibnamefont {Smith}}, \bibinfo
  {author} {\bibfnamefont {M.~G.}\ \bibnamefont {Borselli}}, \bibinfo {author}
  {\bibfnamefont {A.~A.}\ \bibnamefont {Kiselev}}, \bibinfo {author}
  {\bibfnamefont {B.~H.}\ \bibnamefont {Fong}}, \bibinfo {author}
  {\bibfnamefont {K.~S.}\ \bibnamefont {Holabird}}, \bibinfo {author}
  {\bibfnamefont {T.~M.}\ \bibnamefont {Hazard}}, \bibinfo {author}
  {\bibfnamefont {B.}~\bibnamefont {Huang}}, \bibinfo {author} {\bibfnamefont
  {P.~W.}\ \bibnamefont {Deelman}}, \bibinfo {author} {\bibfnamefont
  {I.}~\bibnamefont {Milosavljevic}}, \bibinfo {author} {\bibfnamefont {A.~E.}\
  \bibnamefont {Schmitz}}, \bibinfo {author} {\bibfnamefont {R.~S.}\
  \bibnamefont {Ross}}, \bibinfo {author} {\bibfnamefont {M.~F.}\ \bibnamefont
  {Gyure}},\ and\ \bibinfo {author} {\bibfnamefont {A.~T.}\ \bibnamefont
  {Hunter}},\ }\bibfield  {journal} {\bibinfo  {journal} {Science Advances}\
  }\textbf {\bibinfo {volume} {1}},\ \href
  {https://doi.org/10.1126/sciadv.1500214} {10.1126/sciadv.1500214} (\bibinfo
  {year} {2015})\BibitemShut {NoStop}%
\bibitem [{\citenamefont {Muhonen}\ \emph {et~al.}(2015)\citenamefont
  {Muhonen}, \citenamefont {Laucht}, \citenamefont {Simmons}, \citenamefont
  {Dehollain}, \citenamefont {Kalra}, \citenamefont {Hudson}, \citenamefont
  {Freer}, \citenamefont {Itoh}, \citenamefont {Jamieson}, \citenamefont
  {McCallum}, \citenamefont {Dzurak},\ and\ \citenamefont
  {Morello}}]{Muhonen2015}%
  \BibitemOpen
  \bibfield  {author} {\bibinfo {author} {\bibfnamefont {J.~T.}\ \bibnamefont
  {Muhonen}}, \bibinfo {author} {\bibfnamefont {A.}~\bibnamefont {Laucht}},
  \bibinfo {author} {\bibfnamefont {S.}~\bibnamefont {Simmons}}, \bibinfo
  {author} {\bibfnamefont {J.~P.}\ \bibnamefont {Dehollain}}, \bibinfo {author}
  {\bibfnamefont {R.}~\bibnamefont {Kalra}}, \bibinfo {author} {\bibfnamefont
  {F.~E.}\ \bibnamefont {Hudson}}, \bibinfo {author} {\bibfnamefont
  {S.}~\bibnamefont {Freer}}, \bibinfo {author} {\bibfnamefont {K.~M.}\
  \bibnamefont {Itoh}}, \bibinfo {author} {\bibfnamefont {D.~N.}\ \bibnamefont
  {Jamieson}}, \bibinfo {author} {\bibfnamefont {J.~C.}\ \bibnamefont
  {McCallum}}, \bibinfo {author} {\bibfnamefont {A.~S.}\ \bibnamefont
  {Dzurak}},\ and\ \bibinfo {author} {\bibfnamefont {A.}~\bibnamefont
  {Morello}},\ }\href {http://stacks.iop.org/0953-8984/27/i=15/a=154205}
  {\bibfield  {journal} {\bibinfo  {journal} {Journal of Physics: Condensed
  Matter}\ }\textbf {\bibinfo {volume} {27}},\ \bibinfo {pages} {154205}
  (\bibinfo {year} {2015})}\BibitemShut {NoStop}%
\bibitem [{\citenamefont {Kawakami}\ \emph {et~al.}(2016)\citenamefont
  {Kawakami}, \citenamefont {Jullien}, \citenamefont {Scarlino}, \citenamefont
  {Ward}, \citenamefont {Savage}, \citenamefont {Lagally}, \citenamefont
  {Dobrovitski}, \citenamefont {Friesen}, \citenamefont {Coppersmith},
  \citenamefont {Eriksson},\ and\ \citenamefont {Vandersypen}}]{kawakami2016}%
  \BibitemOpen
  \bibfield  {author} {\bibinfo {author} {\bibfnamefont {E.}~\bibnamefont
  {Kawakami}}, \bibinfo {author} {\bibfnamefont {T.}~\bibnamefont {Jullien}},
  \bibinfo {author} {\bibfnamefont {P.}~\bibnamefont {Scarlino}}, \bibinfo
  {author} {\bibfnamefont {D.~R.}\ \bibnamefont {Ward}}, \bibinfo {author}
  {\bibfnamefont {D.~E.}\ \bibnamefont {Savage}}, \bibinfo {author}
  {\bibfnamefont {M.~G.}\ \bibnamefont {Lagally}}, \bibinfo {author}
  {\bibfnamefont {V.~V.}\ \bibnamefont {Dobrovitski}}, \bibinfo {author}
  {\bibfnamefont {M.}~\bibnamefont {Friesen}}, \bibinfo {author} {\bibfnamefont
  {S.~N.}\ \bibnamefont {Coppersmith}}, \bibinfo {author} {\bibfnamefont
  {M.~A.}\ \bibnamefont {Eriksson}},\ and\ \bibinfo {author} {\bibfnamefont
  {L.~M.~K.}\ \bibnamefont {Vandersypen}},\ }\href
  {https://doi.org/10.1073/pnas.1603251113} {\bibfield  {journal} {\bibinfo
  {journal} {Proceedings of the National Academy of Sciences}\ }\textbf
  {\bibinfo {volume} {113}},\ \bibinfo {pages} {11738} (\bibinfo {year}
  {2016})},\ \Eprint
  {https://arxiv.org/abs/http://www.pnas.org/content/113/42/11738.full.pdf}
  {http://www.pnas.org/content/113/42/11738.full.pdf} \BibitemShut {NoStop}%
\bibitem [{\citenamefont {{Andrews}}\ \emph {et~al.}(2019)\citenamefont
  {{Andrews}}, \citenamefont {{Jones}}, \citenamefont {{Reed}}, \citenamefont
  {{Jones}}, \citenamefont {{Ha}}, \citenamefont {{Jura}}, \citenamefont
  {{Kerckhoff}}, \citenamefont {{Levendorf}}, \citenamefont {{Meenehan}},
  \citenamefont {{Merkel}}, \citenamefont {{Smith}}, \citenamefont {{Sun}},
  \citenamefont {{Weinstein}}, \citenamefont {{Rakher}}, \citenamefont
  {{Ladd}},\ and\ \citenamefont {{Borselli}}}]{HRLAndrews2019}%
  \BibitemOpen
  \bibfield  {author} {\bibinfo {author} {\bibfnamefont {R.~W.}\ \bibnamefont
  {{Andrews}}}, \bibinfo {author} {\bibfnamefont {C.}~\bibnamefont {{Jones}}},
  \bibinfo {author} {\bibfnamefont {M.~D.}\ \bibnamefont {{Reed}}}, \bibinfo
  {author} {\bibfnamefont {A.~M.}\ \bibnamefont {{Jones}}}, \bibinfo {author}
  {\bibfnamefont {S.~D.}\ \bibnamefont {{Ha}}}, \bibinfo {author}
  {\bibfnamefont {M.~P.}\ \bibnamefont {{Jura}}}, \bibinfo {author}
  {\bibfnamefont {J.}~\bibnamefont {{Kerckhoff}}}, \bibinfo {author}
  {\bibfnamefont {M.}~\bibnamefont {{Levendorf}}}, \bibinfo {author}
  {\bibfnamefont {S.}~\bibnamefont {{Meenehan}}}, \bibinfo {author}
  {\bibfnamefont {S.~T.}\ \bibnamefont {{Merkel}}}, \bibinfo {author}
  {\bibfnamefont {A.}~\bibnamefont {{Smith}}}, \bibinfo {author} {\bibfnamefont
  {B.}~\bibnamefont {{Sun}}}, \bibinfo {author} {\bibfnamefont {A.~J.}\
  \bibnamefont {{Weinstein}}}, \bibinfo {author} {\bibfnamefont {M.~T.}\
  \bibnamefont {{Rakher}}}, \bibinfo {author} {\bibfnamefont {T.~D.}\
  \bibnamefont {{Ladd}}},\ and\ \bibinfo {author} {\bibfnamefont {M.~G.}\
  \bibnamefont {{Borselli}}},\ }\href@noop {} {\bibfield  {journal} {\bibinfo
  {journal} {Nature nanotechnology}\ }\textbf {\bibinfo {volume} {14}},\
  \bibinfo {pages} {747} (\bibinfo {year} {2019})}\BibitemShut {NoStop}%
\bibitem [{\citenamefont {Veldhorst}\ \emph {et~al.}(2015)\citenamefont
  {Veldhorst}, \citenamefont {Yang}, \citenamefont {Hwang}, \citenamefont
  {Huang}, \citenamefont {Dehollain}, \citenamefont {Muhonen}, \citenamefont
  {Simmons}, \citenamefont {Laucht}, \citenamefont {Hudson}, \citenamefont
  {Itoh}, \citenamefont {Morello},\ and\ \citenamefont
  {Dzurak}}]{Veldhorst2015}%
  \BibitemOpen
  \bibfield  {author} {\bibinfo {author} {\bibfnamefont {M.}~\bibnamefont
  {Veldhorst}}, \bibinfo {author} {\bibfnamefont {C.~H.}\ \bibnamefont {Yang}},
  \bibinfo {author} {\bibfnamefont {J.~C.~C.}\ \bibnamefont {Hwang}}, \bibinfo
  {author} {\bibfnamefont {W.}~\bibnamefont {Huang}}, \bibinfo {author}
  {\bibfnamefont {J.~P.}\ \bibnamefont {Dehollain}}, \bibinfo {author}
  {\bibfnamefont {J.~T.}\ \bibnamefont {Muhonen}}, \bibinfo {author}
  {\bibfnamefont {S.}~\bibnamefont {Simmons}}, \bibinfo {author} {\bibfnamefont
  {A.}~\bibnamefont {Laucht}}, \bibinfo {author} {\bibfnamefont {F.~E.}\
  \bibnamefont {Hudson}}, \bibinfo {author} {\bibfnamefont {K.~M.}\
  \bibnamefont {Itoh}}, \bibinfo {author} {\bibfnamefont {A.}~\bibnamefont
  {Morello}},\ and\ \bibinfo {author} {\bibfnamefont {A.~S.}\ \bibnamefont
  {Dzurak}},\ }\href {http://dx.doi.org/10.1038/nature15263} {\bibfield
  {journal} {\bibinfo  {journal} {Nature}\ }\textbf {\bibinfo {volume} {526}},\
  \bibinfo {pages} {410} (\bibinfo {year} {2015})}\BibitemShut {NoStop}%
\bibitem [{\citenamefont {Zajac}\ \emph {et~al.}(2016)\citenamefont {Zajac},
  \citenamefont {Hazard}, \citenamefont {Mi}, \citenamefont {Nielsen},\ and\
  \citenamefont {Petta}}]{Zajac2016}%
  \BibitemOpen
  \bibfield  {author} {\bibinfo {author} {\bibfnamefont {D.~M.}\ \bibnamefont
  {Zajac}}, \bibinfo {author} {\bibfnamefont {T.~M.}\ \bibnamefont {Hazard}},
  \bibinfo {author} {\bibfnamefont {X.}~\bibnamefont {Mi}}, \bibinfo {author}
  {\bibfnamefont {E.}~\bibnamefont {Nielsen}},\ and\ \bibinfo {author}
  {\bibfnamefont {J.~R.}\ \bibnamefont {Petta}},\ }\href
  {https://doi.org/10.1103/PhysRevApplied.6.054013} {\bibfield  {journal}
  {\bibinfo  {journal} {Phys. Rev. Applied}\ }\textbf {\bibinfo {volume} {6}},\
  \bibinfo {pages} {054013} (\bibinfo {year} {2016})}\BibitemShut {NoStop}%
\bibitem [{\citenamefont {Homma}\ \emph {et~al.}(1982)\citenamefont {Homma},
  \citenamefont {Yajima},\ and\ \citenamefont {Harada}}]{homma1982liftoff}%
  \BibitemOpen
  \bibfield  {author} {\bibinfo {author} {\bibfnamefont {Y.}~\bibnamefont
  {Homma}}, \bibinfo {author} {\bibfnamefont {A.}~\bibnamefont {Yajima}},\ and\
  \bibinfo {author} {\bibfnamefont {S.}~\bibnamefont {Harada}},\ }\href
  {https://doi.org/10.1109/JSSC.1982.1051707} {\bibfield  {journal} {\bibinfo
  {journal} {IEEE Journal of Solid-State Circuits}\ }\textbf {\bibinfo {volume}
  {17}},\ \bibinfo {pages} {142} (\bibinfo {year} {1982})}\BibitemShut
  {NoStop}%
\bibitem [{\citenamefont {Maurand}\ \emph {et~al.}(2016)\citenamefont
  {Maurand}, \citenamefont {Jehl}, \citenamefont {Kotekar-Patil}, \citenamefont
  {Corna}, \citenamefont {Bohuslavskyi}, \citenamefont {Lavi{\'e}ville},
  \citenamefont {Hutin}, \citenamefont {Barraud}, \citenamefont {Vinet},
  \citenamefont {Sanquer},\ and\ \citenamefont
  {De~Franceschi}}]{maurand2016cmos}%
  \BibitemOpen
  \bibfield  {author} {\bibinfo {author} {\bibfnamefont {R.}~\bibnamefont
  {Maurand}}, \bibinfo {author} {\bibfnamefont {X.}~\bibnamefont {Jehl}},
  \bibinfo {author} {\bibfnamefont {D.}~\bibnamefont {Kotekar-Patil}}, \bibinfo
  {author} {\bibfnamefont {A.}~\bibnamefont {Corna}}, \bibinfo {author}
  {\bibfnamefont {H.}~\bibnamefont {Bohuslavskyi}}, \bibinfo {author}
  {\bibfnamefont {R.}~\bibnamefont {Lavi{\'e}ville}}, \bibinfo {author}
  {\bibfnamefont {L.}~\bibnamefont {Hutin}}, \bibinfo {author} {\bibfnamefont
  {S.}~\bibnamefont {Barraud}}, \bibinfo {author} {\bibfnamefont
  {M.}~\bibnamefont {Vinet}}, \bibinfo {author} {\bibfnamefont
  {M.}~\bibnamefont {Sanquer}},\ and\ \bibinfo {author} {\bibfnamefont
  {S.}~\bibnamefont {De~Franceschi}},\ }\href
  {https://doi.org/10.1038/ncomms13575} {\bibfield  {journal} {\bibinfo
  {journal} {Nature Communications}\ }\textbf {\bibinfo {volume} {7}},\
  \bibinfo {pages} {13575} (\bibinfo {year} {2016})}\BibitemShut {NoStop}%
\bibitem [{\citenamefont {Gilbert}\ \emph {et~al.}(2020)\citenamefont
  {Gilbert}, \citenamefont {Saraiva}, \citenamefont {Lim}, \citenamefont
  {Yang}, \citenamefont {Laucht}, \citenamefont {Bertrand}, \citenamefont
  {Rambal}, \citenamefont {Hutin}, \citenamefont {Escott}, \citenamefont
  {Vinet},\ and\ \citenamefont {Dzurak}}]{gilbert2020twobytwo}%
  \BibitemOpen
  \bibfield  {author} {\bibinfo {author} {\bibfnamefont {W.}~\bibnamefont
  {Gilbert}}, \bibinfo {author} {\bibfnamefont {A.}~\bibnamefont {Saraiva}},
  \bibinfo {author} {\bibfnamefont {W.~H.}\ \bibnamefont {Lim}}, \bibinfo
  {author} {\bibfnamefont {C.~H.}\ \bibnamefont {Yang}}, \bibinfo {author}
  {\bibfnamefont {A.}~\bibnamefont {Laucht}}, \bibinfo {author} {\bibfnamefont
  {B.}~\bibnamefont {Bertrand}}, \bibinfo {author} {\bibfnamefont
  {N.}~\bibnamefont {Rambal}}, \bibinfo {author} {\bibfnamefont
  {L.}~\bibnamefont {Hutin}}, \bibinfo {author} {\bibfnamefont {C.~C.}\
  \bibnamefont {Escott}}, \bibinfo {author} {\bibfnamefont {M.}~\bibnamefont
  {Vinet}},\ and\ \bibinfo {author} {\bibfnamefont {A.~S.}\ \bibnamefont
  {Dzurak}},\ }\href {https://doi.org/10.1021/acs.nanolett.0c02397} {\bibfield
  {journal} {\bibinfo  {journal} {Nano Letters}\ }\textbf {\bibinfo {volume}
  {20}},\ \bibinfo {pages} {7882} (\bibinfo {year} {2020})},\ \bibinfo {note}
  {pMID: 33108202},\ \Eprint
  {https://arxiv.org/abs/https://doi.org/10.1021/acs.nanolett.0c02397}
  {https://doi.org/10.1021/acs.nanolett.0c02397} \BibitemShut {NoStop}%
\bibitem [{\citenamefont {Chanrion}\ \emph {et~al.}(2020)\citenamefont
  {Chanrion}, \citenamefont {Niegemann}, \citenamefont {Bertrand},
  \citenamefont {Spence}, \citenamefont {Jadot}, \citenamefont {Li},
  \citenamefont {Mortemousque}, \citenamefont {Hutin}, \citenamefont {Maurand},
  \citenamefont {Jehl}, \citenamefont {Sanquer}, \citenamefont {De~Franceschi},
  \citenamefont {B\"auerle}, \citenamefont {Balestro}, \citenamefont {Niquet},
  \citenamefont {Vinet}, \citenamefont {Meunier},\ and\ \citenamefont
  {Urdampilleta}}]{chanrion2020twobyfour}%
  \BibitemOpen
  \bibfield  {author} {\bibinfo {author} {\bibfnamefont {E.}~\bibnamefont
  {Chanrion}}, \bibinfo {author} {\bibfnamefont {D.~J.}\ \bibnamefont
  {Niegemann}}, \bibinfo {author} {\bibfnamefont {B.}~\bibnamefont {Bertrand}},
  \bibinfo {author} {\bibfnamefont {C.}~\bibnamefont {Spence}}, \bibinfo
  {author} {\bibfnamefont {B.}~\bibnamefont {Jadot}}, \bibinfo {author}
  {\bibfnamefont {J.}~\bibnamefont {Li}}, \bibinfo {author} {\bibfnamefont
  {P.-A.}\ \bibnamefont {Mortemousque}}, \bibinfo {author} {\bibfnamefont
  {L.}~\bibnamefont {Hutin}}, \bibinfo {author} {\bibfnamefont
  {R.}~\bibnamefont {Maurand}}, \bibinfo {author} {\bibfnamefont
  {X.}~\bibnamefont {Jehl}}, \bibinfo {author} {\bibfnamefont {M.}~\bibnamefont
  {Sanquer}}, \bibinfo {author} {\bibfnamefont {S.}~\bibnamefont
  {De~Franceschi}}, \bibinfo {author} {\bibfnamefont {C.}~\bibnamefont
  {B\"auerle}}, \bibinfo {author} {\bibfnamefont {F.}~\bibnamefont {Balestro}},
  \bibinfo {author} {\bibfnamefont {Y.-M.}\ \bibnamefont {Niquet}}, \bibinfo
  {author} {\bibfnamefont {M.}~\bibnamefont {Vinet}}, \bibinfo {author}
  {\bibfnamefont {T.}~\bibnamefont {Meunier}},\ and\ \bibinfo {author}
  {\bibfnamefont {M.}~\bibnamefont {Urdampilleta}},\ }\href
  {https://doi.org/10.1103/PhysRevApplied.14.024066} {\bibfield  {journal}
  {\bibinfo  {journal} {Phys. Rev. Applied}\ }\textbf {\bibinfo {volume}
  {14}},\ \bibinfo {pages} {024066} (\bibinfo {year} {2020})}\BibitemShut
  {NoStop}%
\bibitem [{\citenamefont {Zwerver}\ \emph {et~al.}(2021)\citenamefont
  {Zwerver}, \citenamefont {Krähenmann}, \citenamefont {Watson}, \citenamefont
  {Lampert}, \citenamefont {George}, \citenamefont {Pillarisetty},
  \citenamefont {Bojarski}, \citenamefont {Amin}, \citenamefont {Amitonov},
  \citenamefont {Boter}, \citenamefont {Caudillo}, \citenamefont
  {Corras-Serrano}, \citenamefont {Dehollain}, \citenamefont {Droulers},
  \citenamefont {Henry}, \citenamefont {Kotlyar}, \citenamefont {Lodari},
  \citenamefont {Luthi}, \citenamefont {Michalak}, \citenamefont {Mueller},
  \citenamefont {Neyens}, \citenamefont {Roberts}, \citenamefont {Samkharadze},
  \citenamefont {Zheng}, \citenamefont {Zietz}, \citenamefont {Scappucci},
  \citenamefont {Veldhorst}, \citenamefont {Vandersypen},\ and\ \citenamefont
  {Clarke}}]{zwerver2021intelqb}%
  \BibitemOpen
  \bibfield  {author} {\bibinfo {author} {\bibfnamefont {A.~M.~J.}\
  \bibnamefont {Zwerver}}, \bibinfo {author} {\bibfnamefont {T.}~\bibnamefont
  {Krähenmann}}, \bibinfo {author} {\bibfnamefont {T.~F.}\ \bibnamefont
  {Watson}}, \bibinfo {author} {\bibfnamefont {L.}~\bibnamefont {Lampert}},
  \bibinfo {author} {\bibfnamefont {H.~C.}\ \bibnamefont {George}}, \bibinfo
  {author} {\bibfnamefont {R.}~\bibnamefont {Pillarisetty}}, \bibinfo {author}
  {\bibfnamefont {S.~A.}\ \bibnamefont {Bojarski}}, \bibinfo {author}
  {\bibfnamefont {P.}~\bibnamefont {Amin}}, \bibinfo {author} {\bibfnamefont
  {S.~V.}\ \bibnamefont {Amitonov}}, \bibinfo {author} {\bibfnamefont {J.~M.}\
  \bibnamefont {Boter}}, \bibinfo {author} {\bibfnamefont {R.}~\bibnamefont
  {Caudillo}}, \bibinfo {author} {\bibfnamefont {D.}~\bibnamefont
  {Corras-Serrano}}, \bibinfo {author} {\bibfnamefont {J.~P.}\ \bibnamefont
  {Dehollain}}, \bibinfo {author} {\bibfnamefont {G.}~\bibnamefont {Droulers}},
  \bibinfo {author} {\bibfnamefont {E.~M.}\ \bibnamefont {Henry}}, \bibinfo
  {author} {\bibfnamefont {R.}~\bibnamefont {Kotlyar}}, \bibinfo {author}
  {\bibfnamefont {M.}~\bibnamefont {Lodari}}, \bibinfo {author} {\bibfnamefont
  {F.}~\bibnamefont {Luthi}}, \bibinfo {author} {\bibfnamefont {D.~J.}\
  \bibnamefont {Michalak}}, \bibinfo {author} {\bibfnamefont {B.~K.}\
  \bibnamefont {Mueller}}, \bibinfo {author} {\bibfnamefont {S.}~\bibnamefont
  {Neyens}}, \bibinfo {author} {\bibfnamefont {J.}~\bibnamefont {Roberts}},
  \bibinfo {author} {\bibfnamefont {N.}~\bibnamefont {Samkharadze}}, \bibinfo
  {author} {\bibfnamefont {G.}~\bibnamefont {Zheng}}, \bibinfo {author}
  {\bibfnamefont {O.~K.}\ \bibnamefont {Zietz}}, \bibinfo {author}
  {\bibfnamefont {G.}~\bibnamefont {Scappucci}}, \bibinfo {author}
  {\bibfnamefont {M.}~\bibnamefont {Veldhorst}}, \bibinfo {author}
  {\bibfnamefont {L.~M.~K.}\ \bibnamefont {Vandersypen}},\ and\ \bibinfo
  {author} {\bibfnamefont {J.~S.}\ \bibnamefont {Clarke}},\ }\Eprint
  {https://arxiv.org/abs/2101.12650} {arXiv:2101.12650 [cond-mat.mes-hall]}
  (\bibinfo {year} {2021})\BibitemShut {NoStop}%
\bibitem [{\citenamefont {Russ}\ and\ \citenamefont
  {Burkard}(2017)}]{russ2016}%
  \BibitemOpen
  \bibfield  {author} {\bibinfo {author} {\bibfnamefont {M.}~\bibnamefont
  {Russ}}\ and\ \bibinfo {author} {\bibfnamefont {G.}~\bibnamefont {Burkard}},\
  }\href {http://stacks.iop.org/0953-8984/29/i=39/a=393001} {\bibfield
  {journal} {\bibinfo  {journal} {Journal of Physics: Condensed Matter}\
  }\textbf {\bibinfo {volume} {29}},\ \bibinfo {pages} {393001} (\bibinfo
  {year} {2017})}\BibitemShut {NoStop}%
\bibitem [{\citenamefont {Jones}\ \emph {et~al.}(2019)\citenamefont {Jones},
  \citenamefont {Pritchett}, \citenamefont {Chen}, \citenamefont {Keating},
  \citenamefont {Andrews}, \citenamefont {Blumoff}, \citenamefont {De~Lorenzo},
  \citenamefont {Eng}, \citenamefont {Ha}, \citenamefont {Kiselev},
  \citenamefont {Meenehan}, \citenamefont {Merkel}, \citenamefont {Wright},
  \citenamefont {Edge}, \citenamefont {Ross}, \citenamefont {Rakher},
  \citenamefont {Borselli},\ and\ \citenamefont {Hunter}}]{HRLJones2019}%
  \BibitemOpen
  \bibfield  {author} {\bibinfo {author} {\bibfnamefont {A.}~\bibnamefont
  {Jones}}, \bibinfo {author} {\bibfnamefont {E.}~\bibnamefont {Pritchett}},
  \bibinfo {author} {\bibfnamefont {E.}~\bibnamefont {Chen}}, \bibinfo {author}
  {\bibfnamefont {T.}~\bibnamefont {Keating}}, \bibinfo {author} {\bibfnamefont
  {R.}~\bibnamefont {Andrews}}, \bibinfo {author} {\bibfnamefont
  {J.}~\bibnamefont {Blumoff}}, \bibinfo {author} {\bibfnamefont
  {L.}~\bibnamefont {De~Lorenzo}}, \bibinfo {author} {\bibfnamefont
  {K.}~\bibnamefont {Eng}}, \bibinfo {author} {\bibfnamefont {S.}~\bibnamefont
  {Ha}}, \bibinfo {author} {\bibfnamefont {A.}~\bibnamefont {Kiselev}},
  \bibinfo {author} {\bibfnamefont {S.}~\bibnamefont {Meenehan}}, \bibinfo
  {author} {\bibfnamefont {S.}~\bibnamefont {Merkel}}, \bibinfo {author}
  {\bibfnamefont {J.}~\bibnamefont {Wright}}, \bibinfo {author} {\bibfnamefont
  {L.}~\bibnamefont {Edge}}, \bibinfo {author} {\bibfnamefont {R.}~\bibnamefont
  {Ross}}, \bibinfo {author} {\bibfnamefont {M.}~\bibnamefont {Rakher}},
  \bibinfo {author} {\bibfnamefont {M.}~\bibnamefont {Borselli}},\ and\
  \bibinfo {author} {\bibfnamefont {A.}~\bibnamefont {Hunter}},\ }\href
  {https://doi.org/10.1103/PhysRevApplied.12.014026} {\bibfield  {journal}
  {\bibinfo  {journal} {Phys. Rev. Applied}\ }\textbf {\bibinfo {volume}
  {12}},\ \bibinfo {pages} {014026} (\bibinfo {year} {2019})}\BibitemShut
  {NoStop}%
\bibitem [{\citenamefont {Willems~van Beveren}\ \emph
  {et~al.}(2010)\citenamefont {Willems~van Beveren}, \citenamefont {Tan},
  \citenamefont {Lai}, \citenamefont {Dzurak},\ and\ \citenamefont
  {Hamilton}}]{vanbeveren2010supplygate}%
  \BibitemOpen
  \bibfield  {author} {\bibinfo {author} {\bibfnamefont {L.~H.}\ \bibnamefont
  {Willems~van Beveren}}, \bibinfo {author} {\bibfnamefont {K.~Y.}\
  \bibnamefont {Tan}}, \bibinfo {author} {\bibfnamefont {N.~S.}\ \bibnamefont
  {Lai}}, \bibinfo {author} {\bibfnamefont {A.~S.}\ \bibnamefont {Dzurak}},\
  and\ \bibinfo {author} {\bibfnamefont {A.~R.}\ \bibnamefont {Hamilton}},\
  }\href {https://doi.org/10.1063/1.3501136} {\bibfield  {journal} {\bibinfo
  {journal} {Applied Physics Letters}\ }\textbf {\bibinfo {volume} {97}},\
  \bibinfo {pages} {152102} (\bibinfo {year} {2010})},\ \Eprint
  {https://arxiv.org/abs/https://doi.org/10.1063/1.3501136}
  {https://doi.org/10.1063/1.3501136} \BibitemShut {NoStop}%
\bibitem [{\citenamefont {DiVincenzo}\ \emph {et~al.}(2000)\citenamefont
  {DiVincenzo}, \citenamefont {Bacon}, \citenamefont {Kempe}, \citenamefont
  {Burkard},\ and\ \citenamefont {Whaley}}]{divincenzo2000}%
  \BibitemOpen
  \bibfield  {author} {\bibinfo {author} {\bibfnamefont {D.~P.}\ \bibnamefont
  {DiVincenzo}}, \bibinfo {author} {\bibfnamefont {D.}~\bibnamefont {Bacon}},
  \bibinfo {author} {\bibfnamefont {J.}~\bibnamefont {Kempe}}, \bibinfo
  {author} {\bibfnamefont {G.}~\bibnamefont {Burkard}},\ and\ \bibinfo {author}
  {\bibfnamefont {K.~B.}\ \bibnamefont {Whaley}},\ }\href
  {http://dx.doi.org/10.1038/35042541} {\bibfield  {journal} {\bibinfo
  {journal} {Nature}\ }\textbf {\bibinfo {volume} {408}},\ \bibinfo {pages}
  {339} (\bibinfo {year} {2000})}\BibitemShut {NoStop}%
\bibitem [{\citenamefont {Lawrie}\ \emph {et~al.}(2020)\citenamefont {Lawrie},
  \citenamefont {Eenink}, \citenamefont {Hendrickx}, \citenamefont {Boter},
  \citenamefont {Petit}, \citenamefont {Amitonov}, \citenamefont {Lodari},
  \citenamefont {Paquelet~Wuetz}, \citenamefont {Volk}, \citenamefont {Philips}
  \emph {et~al.}}]{lawrie2020quantum}%
  \BibitemOpen
  \bibfield  {author} {\bibinfo {author} {\bibfnamefont {W.}~\bibnamefont
  {Lawrie}}, \bibinfo {author} {\bibfnamefont {H.}~\bibnamefont {Eenink}},
  \bibinfo {author} {\bibfnamefont {N.}~\bibnamefont {Hendrickx}}, \bibinfo
  {author} {\bibfnamefont {J.}~\bibnamefont {Boter}}, \bibinfo {author}
  {\bibfnamefont {L.}~\bibnamefont {Petit}}, \bibinfo {author} {\bibfnamefont
  {S.}~\bibnamefont {Amitonov}}, \bibinfo {author} {\bibfnamefont
  {M.}~\bibnamefont {Lodari}}, \bibinfo {author} {\bibfnamefont
  {B.}~\bibnamefont {Paquelet~Wuetz}}, \bibinfo {author} {\bibfnamefont
  {C.}~\bibnamefont {Volk}}, \bibinfo {author} {\bibfnamefont {S.}~\bibnamefont
  {Philips}}, \emph {et~al.},\ }\href@noop {} {\bibfield  {journal} {\bibinfo
  {journal} {Applied Physics Letters}\ }\textbf {\bibinfo {volume} {116}},\
  \bibinfo {pages} {080501} (\bibinfo {year} {2020})}\BibitemShut {NoStop}%
\bibitem [{\citenamefont {Connors}\ \emph {et~al.}(2019)\citenamefont
  {Connors}, \citenamefont {Nelson}, \citenamefont {Qiao}, \citenamefont
  {Edge},\ and\ \citenamefont {Nichol}}]{connors2019noise}%
  \BibitemOpen
  \bibfield  {author} {\bibinfo {author} {\bibfnamefont {E.~J.}\ \bibnamefont
  {Connors}}, \bibinfo {author} {\bibfnamefont {J.}~\bibnamefont {Nelson}},
  \bibinfo {author} {\bibfnamefont {H.}~\bibnamefont {Qiao}}, \bibinfo {author}
  {\bibfnamefont {L.~F.}\ \bibnamefont {Edge}},\ and\ \bibinfo {author}
  {\bibfnamefont {J.~M.}\ \bibnamefont {Nichol}},\ }\href
  {https://doi.org/10.1103/PhysRevB.100.165305} {\bibfield  {journal} {\bibinfo
   {journal} {Phys. Rev. B}\ }\textbf {\bibinfo {volume} {100}},\ \bibinfo
  {pages} {165305} (\bibinfo {year} {2019})}\BibitemShut {NoStop}%
\bibitem [{\citenamefont {Hauptmann}\ \emph {et~al.}(2009)\citenamefont
  {Hauptmann}, \citenamefont {Choi}, \citenamefont {Jaschinsky}, \citenamefont
  {Hohle}, \citenamefont {Kretz},\ and\ \citenamefont
  {Eng}}]{hauptmann2009pec}%
  \BibitemOpen
  \bibfield  {author} {\bibinfo {author} {\bibfnamefont {M.}~\bibnamefont
  {Hauptmann}}, \bibinfo {author} {\bibfnamefont {K.-H.}\ \bibnamefont {Choi}},
  \bibinfo {author} {\bibfnamefont {P.}~\bibnamefont {Jaschinsky}}, \bibinfo
  {author} {\bibfnamefont {C.}~\bibnamefont {Hohle}}, \bibinfo {author}
  {\bibfnamefont {J.}~\bibnamefont {Kretz}},\ and\ \bibinfo {author}
  {\bibfnamefont {L.}~\bibnamefont {Eng}},\ }\href
  {https://doi.org/https://doi.org/10.1016/j.mee.2008.12.053} {\bibfield
  {journal} {\bibinfo  {journal} {Microelectronic Engineering}\ }\textbf
  {\bibinfo {volume} {86}},\ \bibinfo {pages} {539} (\bibinfo {year} {2009})},\
  \bibinfo {note} {mNE ’08}\BibitemShut {NoStop}%
\bibitem [{\citenamefont {Mi}\ \emph {et~al.}(2017)\citenamefont {Mi},
  \citenamefont {Cady}, \citenamefont {Zajac}, \citenamefont {Stehlik},
  \citenamefont {Edge},\ and\ \citenamefont {Petta}}]{mi2017cqed}%
  \BibitemOpen
  \bibfield  {author} {\bibinfo {author} {\bibfnamefont {X.}~\bibnamefont
  {Mi}}, \bibinfo {author} {\bibfnamefont {J.~V.}\ \bibnamefont {Cady}},
  \bibinfo {author} {\bibfnamefont {D.~M.}\ \bibnamefont {Zajac}}, \bibinfo
  {author} {\bibfnamefont {J.}~\bibnamefont {Stehlik}}, \bibinfo {author}
  {\bibfnamefont {L.~F.}\ \bibnamefont {Edge}},\ and\ \bibinfo {author}
  {\bibfnamefont {J.~R.}\ \bibnamefont {Petta}},\ }\href
  {https://doi.org/10.1063/1.4974536} {\bibfield  {journal} {\bibinfo
  {journal} {Applied Physics Letters}\ }\textbf {\bibinfo {volume} {110}},\
  \bibinfo {pages} {043502} (\bibinfo {year} {2017})},\ \Eprint
  {https://arxiv.org/abs/https://doi.org/10.1063/1.4974536}
  {https://doi.org/10.1063/1.4974536} \BibitemShut {NoStop}%
\bibitem [{\citenamefont {Eenink}\ \emph {et~al.}(2019)\citenamefont {Eenink},
  \citenamefont {Petit}, \citenamefont {Lawrie}, \citenamefont {Clarke},
  \citenamefont {Vandersypen},\ and\ \citenamefont
  {Veldhorst}}]{eenink2019mos}%
  \BibitemOpen
  \bibfield  {author} {\bibinfo {author} {\bibfnamefont {H.~G.~J.}\
  \bibnamefont {Eenink}}, \bibinfo {author} {\bibfnamefont {L.}~\bibnamefont
  {Petit}}, \bibinfo {author} {\bibfnamefont {W.~I.~L.}\ \bibnamefont
  {Lawrie}}, \bibinfo {author} {\bibfnamefont {J.~S.}\ \bibnamefont {Clarke}},
  \bibinfo {author} {\bibfnamefont {L.~M.~K.}\ \bibnamefont {Vandersypen}},\
  and\ \bibinfo {author} {\bibfnamefont {M.}~\bibnamefont {Veldhorst}},\ }\href
  {https://doi.org/10.1021/acs.nanolett.9b03254} {\bibfield  {journal}
  {\bibinfo  {journal} {Nano Letters}\ }\textbf {\bibinfo {volume} {19}},\
  \bibinfo {pages} {8653} (\bibinfo {year} {2019})},\ \bibinfo {note} {pMID:
  31755273},\ \Eprint
  {https://arxiv.org/abs/https://doi.org/10.1021/acs.nanolett.9b03254}
  {https://doi.org/10.1021/acs.nanolett.9b03254} \BibitemShut {NoStop}%
\bibitem [{\citenamefont {Brauns}\ \emph {et~al.}(2018)\citenamefont {Brauns},
  \citenamefont {Amitonov}, \citenamefont {Spruijtenburg},\ and\ \citenamefont
  {Zwanenburg}}]{brauns2018pd}%
  \BibitemOpen
  \bibfield  {author} {\bibinfo {author} {\bibfnamefont {M.}~\bibnamefont
  {Brauns}}, \bibinfo {author} {\bibfnamefont {S.~V.}\ \bibnamefont
  {Amitonov}}, \bibinfo {author} {\bibfnamefont {P.-C.}\ \bibnamefont
  {Spruijtenburg}},\ and\ \bibinfo {author} {\bibfnamefont {F.~A.}\
  \bibnamefont {Zwanenburg}},\ }\href
  {https://doi.org/10.1038/s41598-018-24004-y} {\bibfield  {journal} {\bibinfo
  {journal} {Scientific Reports}\ }\textbf {\bibinfo {volume} {8}},\ \bibinfo
  {pages} {5690} (\bibinfo {year} {2018})}\BibitemShut {NoStop}%
\bibitem [{\citenamefont {Efros}\ \emph {et~al.}(1993)\citenamefont {Efros},
  \citenamefont {Pikus},\ and\ \citenamefont {Burnett}}]{efros1993}%
  \BibitemOpen
  \bibfield  {author} {\bibinfo {author} {\bibfnamefont {A.~L.}\ \bibnamefont
  {Efros}}, \bibinfo {author} {\bibfnamefont {F.~G.}\ \bibnamefont {Pikus}},\
  and\ \bibinfo {author} {\bibfnamefont {V.~G.}\ \bibnamefont {Burnett}},\
  }\href {https://doi.org/10.1103/PhysRevB.47.2233} {\bibfield  {journal}
  {\bibinfo  {journal} {Phys. Rev. B}\ }\textbf {\bibinfo {volume} {47}},\
  \bibinfo {pages} {2233} (\bibinfo {year} {1993})}\BibitemShut {NoStop}%
\bibitem [{\citenamefont {Das~Sarma}\ and\ \citenamefont
  {Hwang}(2014)}]{dassarma2014}%
  \BibitemOpen
  \bibfield  {author} {\bibinfo {author} {\bibfnamefont {S.}~\bibnamefont
  {Das~Sarma}}\ and\ \bibinfo {author} {\bibfnamefont {E.~H.}\ \bibnamefont
  {Hwang}},\ }\href {https://doi.org/10.1103/PhysRevB.89.235423} {\bibfield
  {journal} {\bibinfo  {journal} {Phys. Rev. B}\ }\textbf {\bibinfo {volume}
  {89}},\ \bibinfo {pages} {235423} (\bibinfo {year} {2014})}\BibitemShut
  {NoStop}%
\bibitem [{\citenamefont {Davies}\ \emph {et~al.}(1995)\citenamefont {Davies},
  \citenamefont {Larkin},\ and\ \citenamefont {Sukhorukov}}]{davies1995}%
  \BibitemOpen
  \bibfield  {author} {\bibinfo {author} {\bibfnamefont {J.~H.}\ \bibnamefont
  {Davies}}, \bibinfo {author} {\bibfnamefont {I.~A.}\ \bibnamefont {Larkin}},\
  and\ \bibinfo {author} {\bibfnamefont {E.~V.}\ \bibnamefont {Sukhorukov}},\
  }\href {https://doi.org/10.1063/1.359446} {\bibfield  {journal} {\bibinfo
  {journal} {Journal of Applied Physics}\ }\textbf {\bibinfo {volume} {77}},\
  \bibinfo {pages} {4504} (\bibinfo {year} {1995})},\ \Eprint
  {https://arxiv.org/abs/https://doi.org/10.1063/1.359446}
  {https://doi.org/10.1063/1.359446} \BibitemShut {NoStop}%
\bibitem [{\citenamefont {Borselli}\ \emph {et~al.}(2015)\citenamefont
  {Borselli}, \citenamefont {Eng}, \citenamefont {Ross}, \citenamefont
  {Hazard}, \citenamefont {Holabird}, \citenamefont {Huang}, \citenamefont
  {Kiselev}, \citenamefont {Deelman}, \citenamefont {Warren}, \citenamefont
  {Milosavljevic}, \citenamefont {Schmitz}, \citenamefont {Sokolich},
  \citenamefont {Gyure},\ and\ \citenamefont {Hunter}}]{Borselli2015}%
  \BibitemOpen
  \bibfield  {author} {\bibinfo {author} {\bibfnamefont {M.~G.}\ \bibnamefont
  {Borselli}}, \bibinfo {author} {\bibfnamefont {K.}~\bibnamefont {Eng}},
  \bibinfo {author} {\bibfnamefont {R.~S.}\ \bibnamefont {Ross}}, \bibinfo
  {author} {\bibfnamefont {T.~M.}\ \bibnamefont {Hazard}}, \bibinfo {author}
  {\bibfnamefont {K.~S.}\ \bibnamefont {Holabird}}, \bibinfo {author}
  {\bibfnamefont {B.}~\bibnamefont {Huang}}, \bibinfo {author} {\bibfnamefont
  {A.~A.}\ \bibnamefont {Kiselev}}, \bibinfo {author} {\bibfnamefont {P.~W.}\
  \bibnamefont {Deelman}}, \bibinfo {author} {\bibfnamefont {L.~D.}\
  \bibnamefont {Warren}}, \bibinfo {author} {\bibfnamefont {I.}~\bibnamefont
  {Milosavljevic}}, \bibinfo {author} {\bibfnamefont {A.~E.}\ \bibnamefont
  {Schmitz}}, \bibinfo {author} {\bibfnamefont {M.}~\bibnamefont {Sokolich}},
  \bibinfo {author} {\bibfnamefont {M.~F.}\ \bibnamefont {Gyure}},\ and\
  \bibinfo {author} {\bibfnamefont {A.~T.}\ \bibnamefont {Hunter}},\ }\href
  {http://stacks.iop.org/0957-4484/26/i=37/a=375202} {\bibfield  {journal}
  {\bibinfo  {journal} {Nanotechnology}\ }\textbf {\bibinfo {volume} {26}},\
  \bibinfo {pages} {375202} (\bibinfo {year} {2015})}\BibitemShut {NoStop}%
\bibitem [{\citenamefont {Xue}\ \emph {et~al.}(2020)\citenamefont {Xue},
  \citenamefont {Patra}, \citenamefont {van Dijk}, \citenamefont {Samkharadze},
  \citenamefont {Subramanian}, \citenamefont {Corna}, \citenamefont {Jeon},
  \citenamefont {Sheikh}, \citenamefont {Juarez-Hernandez}, \citenamefont
  {Esparza}, \citenamefont {Rampurawala}, \citenamefont {Carlton},
  \citenamefont {Ravikumar}, \citenamefont {Nieva}, \citenamefont {Kim},
  \citenamefont {Lee}, \citenamefont {Sammak}, \citenamefont {Scappucci},
  \citenamefont {Veldhorst}, \citenamefont {Sebastiano}, \citenamefont
  {Babaie}, \citenamefont {Pellerano}, \citenamefont {Charbon},\ and\
  \citenamefont {Vandersypen}}]{xue2020horseridge}%
  \BibitemOpen
  \bibfield  {author} {\bibinfo {author} {\bibfnamefont {X.}~\bibnamefont
  {Xue}}, \bibinfo {author} {\bibfnamefont {B.}~\bibnamefont {Patra}}, \bibinfo
  {author} {\bibfnamefont {J.~P.~G.}\ \bibnamefont {van Dijk}}, \bibinfo
  {author} {\bibfnamefont {N.}~\bibnamefont {Samkharadze}}, \bibinfo {author}
  {\bibfnamefont {S.}~\bibnamefont {Subramanian}}, \bibinfo {author}
  {\bibfnamefont {A.}~\bibnamefont {Corna}}, \bibinfo {author} {\bibfnamefont
  {C.}~\bibnamefont {Jeon}}, \bibinfo {author} {\bibfnamefont {F.}~\bibnamefont
  {Sheikh}}, \bibinfo {author} {\bibfnamefont {E.}~\bibnamefont
  {Juarez-Hernandez}}, \bibinfo {author} {\bibfnamefont {B.~P.}\ \bibnamefont
  {Esparza}}, \bibinfo {author} {\bibfnamefont {H.}~\bibnamefont
  {Rampurawala}}, \bibinfo {author} {\bibfnamefont {B.}~\bibnamefont
  {Carlton}}, \bibinfo {author} {\bibfnamefont {S.}~\bibnamefont {Ravikumar}},
  \bibinfo {author} {\bibfnamefont {C.}~\bibnamefont {Nieva}}, \bibinfo
  {author} {\bibfnamefont {S.}~\bibnamefont {Kim}}, \bibinfo {author}
  {\bibfnamefont {H.-J.}\ \bibnamefont {Lee}}, \bibinfo {author} {\bibfnamefont
  {A.}~\bibnamefont {Sammak}}, \bibinfo {author} {\bibfnamefont
  {G.}~\bibnamefont {Scappucci}}, \bibinfo {author} {\bibfnamefont
  {M.}~\bibnamefont {Veldhorst}}, \bibinfo {author} {\bibfnamefont
  {F.}~\bibnamefont {Sebastiano}}, \bibinfo {author} {\bibfnamefont
  {M.}~\bibnamefont {Babaie}}, \bibinfo {author} {\bibfnamefont
  {S.}~\bibnamefont {Pellerano}}, \bibinfo {author} {\bibfnamefont
  {E.}~\bibnamefont {Charbon}},\ and\ \bibinfo {author} {\bibfnamefont
  {L.~M.~K.}\ \bibnamefont {Vandersypen}},\ }\Eprint
  {https://arxiv.org/abs/2009.14185} {arXiv:2009.14185 [quant-ph]}  (\bibinfo
  {year} {2020})\BibitemShut {NoStop}%
\bibitem [{\citenamefont {Reed}\ \emph {et~al.}(2016)\citenamefont {Reed},
  \citenamefont {Maune}, \citenamefont {Andrews}, \citenamefont {Borselli},
  \citenamefont {Eng}, \citenamefont {Jura}, \citenamefont {Kiselev},
  \citenamefont {Ladd}, \citenamefont {Merkel}, \citenamefont {Milosavljevic},
  \citenamefont {Pritchett}, \citenamefont {Rakher}, \citenamefont {Ross},
  \citenamefont {Schmitz}, \citenamefont {Smith}, \citenamefont {Wright},
  \citenamefont {Gyure},\ and\ \citenamefont {Hunter}}]{reed2016}%
  \BibitemOpen
  \bibfield  {author} {\bibinfo {author} {\bibfnamefont {M.~D.}\ \bibnamefont
  {Reed}}, \bibinfo {author} {\bibfnamefont {B.~M.}\ \bibnamefont {Maune}},
  \bibinfo {author} {\bibfnamefont {R.~W.}\ \bibnamefont {Andrews}}, \bibinfo
  {author} {\bibfnamefont {M.~G.}\ \bibnamefont {Borselli}}, \bibinfo {author}
  {\bibfnamefont {K.}~\bibnamefont {Eng}}, \bibinfo {author} {\bibfnamefont
  {M.~P.}\ \bibnamefont {Jura}}, \bibinfo {author} {\bibfnamefont {A.~A.}\
  \bibnamefont {Kiselev}}, \bibinfo {author} {\bibfnamefont {T.~D.}\
  \bibnamefont {Ladd}}, \bibinfo {author} {\bibfnamefont {S.~T.}\ \bibnamefont
  {Merkel}}, \bibinfo {author} {\bibfnamefont {I.}~\bibnamefont
  {Milosavljevic}}, \bibinfo {author} {\bibfnamefont {E.~J.}\ \bibnamefont
  {Pritchett}}, \bibinfo {author} {\bibfnamefont {M.~T.}\ \bibnamefont
  {Rakher}}, \bibinfo {author} {\bibfnamefont {R.~S.}\ \bibnamefont {Ross}},
  \bibinfo {author} {\bibfnamefont {A.~E.}\ \bibnamefont {Schmitz}}, \bibinfo
  {author} {\bibfnamefont {A.}~\bibnamefont {Smith}}, \bibinfo {author}
  {\bibfnamefont {J.~A.}\ \bibnamefont {Wright}}, \bibinfo {author}
  {\bibfnamefont {M.~F.}\ \bibnamefont {Gyure}},\ and\ \bibinfo {author}
  {\bibfnamefont {A.~T.}\ \bibnamefont {Hunter}},\ }\href
  {https://doi.org/10.1103/PhysRevLett.116.110402} {\bibfield  {journal}
  {\bibinfo  {journal} {Phys. Rev. Lett.}\ }\textbf {\bibinfo {volume} {116}},\
  \bibinfo {pages} {110402} (\bibinfo {year} {2016})}\BibitemShut {NoStop}%
\end{thebibliography}%


\begin{thebibliography}{2}%
\makeatletter
\providecommand \@ifxundefined [1]{%
 \@ifx{#1\undefined}
}%
\providecommand \@ifnum [1]{%
 \ifnum #1\expandafter \@firstoftwo
 \else \expandafter \@secondoftwo
 \fi
}%
\providecommand \@ifx [1]{%
 \ifx #1\expandafter \@firstoftwo
 \else \expandafter \@secondoftwo
 \fi
}%
\providecommand \natexlab [1]{#1}%
\providecommand \enquote  [1]{``#1''}%
\providecommand \bibnamefont  [1]{#1}%
\providecommand \bibfnamefont [1]{#1}%
\providecommand \citenamefont [1]{#1}%
\providecommand \href@noop [0]{\@secondoftwo}%
\providecommand \href [0]{\begingroup \@sanitize@url \@href}%
\providecommand \@href[1]{\@@startlink{#1}\@@href}%
\providecommand \@@href[1]{\endgroup#1\@@endlink}%
\providecommand \@sanitize@url [0]{\catcode `\\12\catcode `\$12\catcode
  `\&12\catcode `\#12\catcode `\^12\catcode `\_12\catcode `\%12\relax}%
\providecommand \@@startlink[1]{}%
\providecommand \@@endlink[0]{}%
\providecommand \url  [0]{\begingroup\@sanitize@url \@url }%
\providecommand \@url [1]{\endgroup\@href {#1}{\urlprefix }}%
\providecommand \urlprefix  [0]{URL }%
\providecommand \Eprint [0]{\href }%
\providecommand \doibase [0]{https://doi.org/}%
\providecommand \selectlanguage [0]{\@gobble}%
\providecommand \bibinfo  [0]{\@secondoftwo}%
\providecommand \bibfield  [0]{\@secondoftwo}%
\providecommand \translation [1]{[#1]}%
\providecommand \BibitemOpen [0]{}%
\providecommand \bibitemStop [0]{}%
\providecommand \bibitemNoStop [0]{.\EOS\space}%
\providecommand \EOS [0]{\spacefactor3000\relax}%
\providecommand \BibitemShut  [1]{\csname bibitem#1\endcsname}%
\let\auto@bib@innerbib\@empty
%</preamble>
\bibitem [{\citenamefont {Zajac}\ \emph {et~al.}(2018)\citenamefont {Zajac},
  \citenamefont {Sigillito}, \citenamefont {Russ}, \citenamefont {Borjans},
  \citenamefont {Taylor}, \citenamefont {Burkard},\ and\ \citenamefont
  {Petta}}]{Zajac2018}%
  \BibitemOpen
  \bibfield  {author} {\bibinfo {author} {\bibfnamefont {D.~M.}\ \bibnamefont
  {Zajac}}, \bibinfo {author} {\bibfnamefont {A.~J.}\ \bibnamefont
  {Sigillito}}, \bibinfo {author} {\bibfnamefont {M.}~\bibnamefont {Russ}},
  \bibinfo {author} {\bibfnamefont {F.}~\bibnamefont {Borjans}}, \bibinfo
  {author} {\bibfnamefont {J.~M.}\ \bibnamefont {Taylor}}, \bibinfo {author}
  {\bibfnamefont {G.}~\bibnamefont {Burkard}},\ and\ \bibinfo {author}
  {\bibfnamefont {J.~R.}\ \bibnamefont {Petta}},\ }\href
  {https://doi.org/10.1126/science.aao5965} {\bibfield  {journal} {\bibinfo
  {journal} {Science}\ }\textbf {\bibinfo {volume} {359}},\ \bibinfo {pages}
  {439} (\bibinfo {year} {2018})}\BibitemShut {NoStop}%
\bibitem [{\citenamefont {Reed}\ \emph {et~al.}(2016)\citenamefont {Reed},
  \citenamefont {Maune}, \citenamefont {Andrews}, \citenamefont {Borselli},
  \citenamefont {Eng}, \citenamefont {Jura}, \citenamefont {Kiselev},
  \citenamefont {Ladd}, \citenamefont {Merkel}, \citenamefont {Milosavljevic},
  \citenamefont {Pritchett}, \citenamefont {Rakher}, \citenamefont {Ross},
  \citenamefont {Schmitz}, \citenamefont {Smith}, \citenamefont {Wright},
  \citenamefont {Gyure},\ and\ \citenamefont {Hunter}}]{reed2016}%
  \BibitemOpen
  \bibfield  {author} {\bibinfo {author} {\bibfnamefont {M.~D.}\ \bibnamefont
  {Reed}}, \bibinfo {author} {\bibfnamefont {B.~M.}\ \bibnamefont {Maune}},
  \bibinfo {author} {\bibfnamefont {R.~W.}\ \bibnamefont {Andrews}}, \bibinfo
  {author} {\bibfnamefont {M.~G.}\ \bibnamefont {Borselli}}, \bibinfo {author}
  {\bibfnamefont {K.}~\bibnamefont {Eng}}, \bibinfo {author} {\bibfnamefont
  {M.~P.}\ \bibnamefont {Jura}}, \bibinfo {author} {\bibfnamefont {A.~A.}\
  \bibnamefont {Kiselev}}, \bibinfo {author} {\bibfnamefont {T.~D.}\
  \bibnamefont {Ladd}}, \bibinfo {author} {\bibfnamefont {S.~T.}\ \bibnamefont
  {Merkel}}, \bibinfo {author} {\bibfnamefont {I.}~\bibnamefont
  {Milosavljevic}}, \bibinfo {author} {\bibfnamefont {E.~J.}\ \bibnamefont
  {Pritchett}}, \bibinfo {author} {\bibfnamefont {M.~T.}\ \bibnamefont
  {Rakher}}, \bibinfo {author} {\bibfnamefont {R.~S.}\ \bibnamefont {Ross}},
  \bibinfo {author} {\bibfnamefont {A.~E.}\ \bibnamefont {Schmitz}}, \bibinfo
  {author} {\bibfnamefont {A.}~\bibnamefont {Smith}}, \bibinfo {author}
  {\bibfnamefont {J.~A.}\ \bibnamefont {Wright}}, \bibinfo {author}
  {\bibfnamefont {M.~F.}\ \bibnamefont {Gyure}},\ and\ \bibinfo {author}
  {\bibfnamefont {A.~T.}\ \bibnamefont {Hunter}},\ }\href
  {https://doi.org/10.1103/PhysRevLett.116.110402} {\bibfield  {journal}
  {\bibinfo  {journal} {Phys. Rev. Lett.}\ }\textbf {\bibinfo {volume} {116}},\
  \bibinfo {pages} {110402} (\bibinfo {year} {2016})}\BibitemShut {NoStop}%
\end{thebibliography}%
\newpage

\end{document}

\author{Wonill Ha}
\author{Sieu D. Ha}
\author{Maxwell D. Choi}
\author{Yan Tang}
\author{Adele E. Schmitz}
\author{Mark P. Levendorf}
\author{Kangmu Lee}
\author{James M. Chappell}
\author{Tower S. Adams}
\author{Daniel R. Hulbert}
\author{Edwin Acuna}
\author{Ramsey S. Noah}
\author{Justine W. Matten}
\author{Michael P. Jura}
\author{Jeffrey A. Wright}
\author{Matthew T. Rakher}
\author{Matthew G. Borselli}
\affiliation{HRL Laboratories, LLC, 3011 Malibu Canyon Road, Malibu, California 90265, USA}

\begin{abstract}
\end{abstract}

\title{A flexible design platform for Si/SiGe exchange-only qubits with low disorder\newline \newline
Supplemental Materials}
\maketitle

\begin{figure}[h]
\includegraphics[width=.7\linewidth]{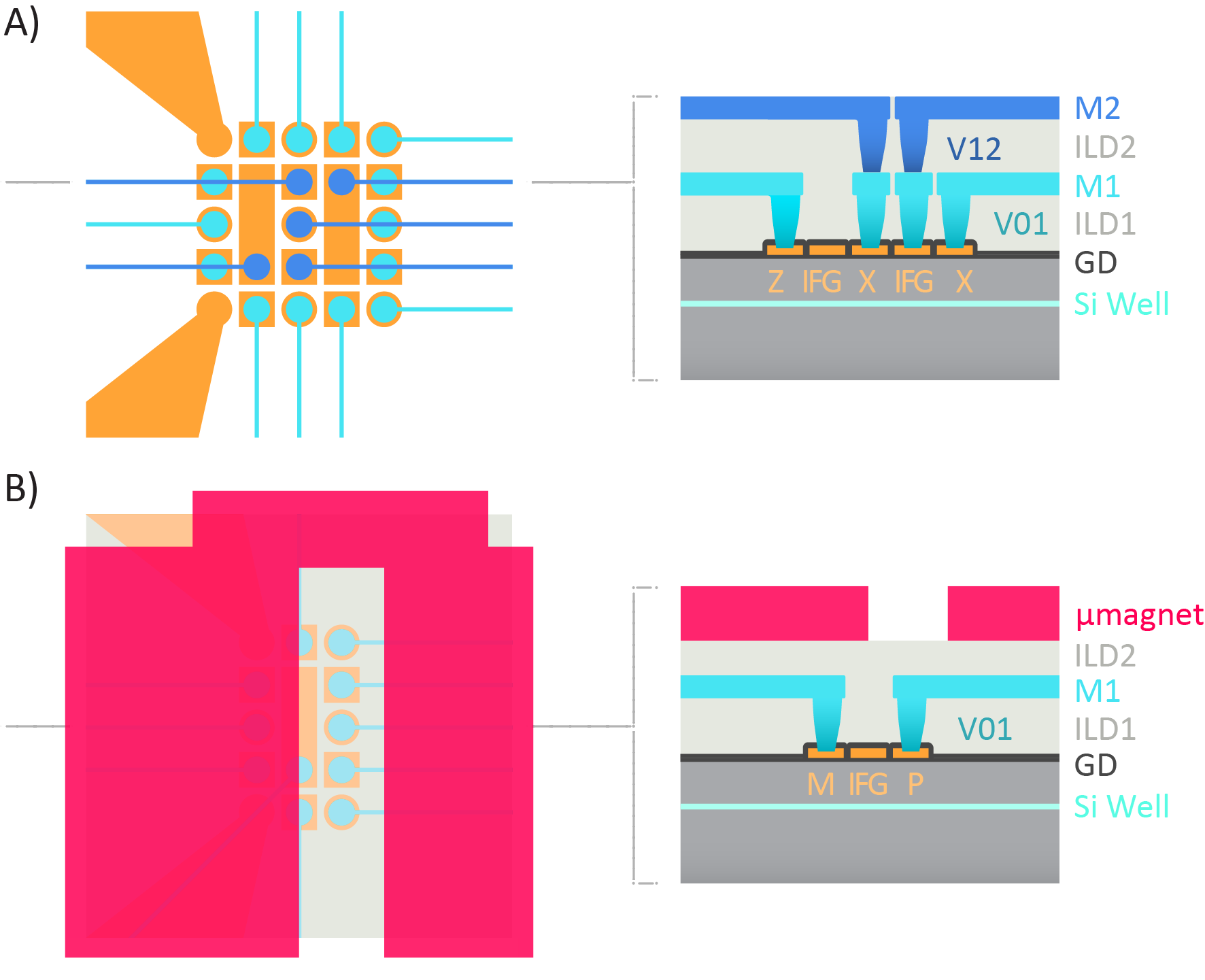}
\caption{(A) Illustrations of a 2x3 dot device connected with 2 BEOL levels. Top-down is left, and dashed lines denote cut of cross-section on right.  In the top-down image, M- and Z-gates are on the left between bath gates, a 3-dot is in the center, and another 3-dot is on the right.  T-gates and IFG islands separate 3-dot columns.  M1 connects to all perimeter gates and M2 connects to all interior gates.  (B) Illustration of a 3-dot device with a micromagnet patterned on top.  Top-down is left, and dashed lines denote cut of cross-section on right. 
}
\label{sfigdev}
\end{figure}

The ability to extend the SLEDGE gate layout and add additional device features is shown in Fig.~\ref{sfigdev}.  In Fig.~\ref{sfigdev}(a) we show how a SLEDGE 2x3 quantum dot device, which is 3 rows of dots including the M- and Z-gates, can be connected by adding an additional BEOL level M2.  The perimeter gates would be connected with M1 while the interior gates would be connected with M2.  The M2 process and pitch could be nominally the same as that of M1, allowing for ease of extending the process flow.  Such designs are enabled by the dot-shaped gates and interconnect process of SLEDGE.  Similarly, we show in Fig.~\ref{sfigdev}(b) how a micromagnet can be straightforwardly fabricated on a 3-dot device after M1 patterning.  The micromagnet is drawn as in~\cite{Zajac2018}.  An additional dielectric layer (ILD2) could be deposited after M1 and the micromagnet could be optically patterned directly on top of ILD2 by lift-off or subtractive etch processes.

\newpage
\begin{figure}[h]
\includegraphics[width=0.6\linewidth]{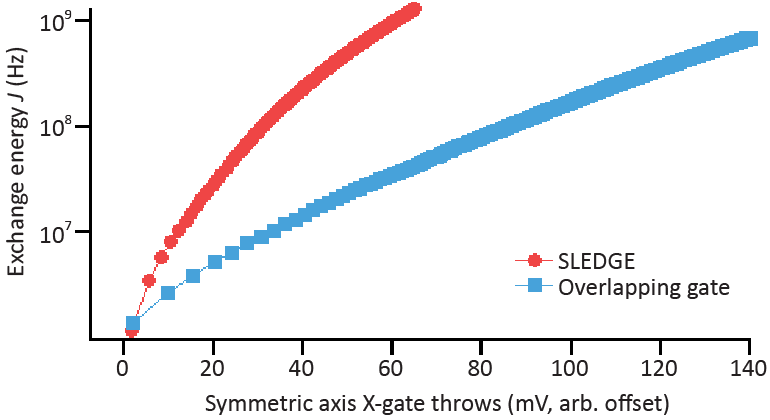}
\centering
\caption{Calibrated exchange energy as a function of X-gate throw on the symmetric axis for SLEDGE gates (red) and overlapping gates (blue).  Curves are offset to overlap at \SI{0}{\mV}.
}
\label{sfigexchange}
\end{figure}

In Fig.~\ref{sfigexchange} we show that uniplanar SLEDGE gates have greater exchange energy ($J$) dynamic range than overlapping gates as a function of X-gate voltage.  The overlapping gates here are patterned by lift-off metallization.  Exchange energy is calibrated by measuring exchange oscillation frequency on the symmetric axis~\cite{reed2016} with fixed voltage pulse duration as a function of X-gate voltage.  It can be seen in the figure that SLEDGE X-gates can achieve higher $J$ than overlapping X-gates for a given voltage throw.  For example, at \SI{40}{\mV} offset, the exchange energy with SLEDGE gates is \num{>10}$\times$ larger than with overlapping gates.  In addition, the exchange energy in the absence of X-gate throw is lower with SLEDGE as compared to overlapping gates, which minimizes unintentional rotations during idle.  The advantages of SLEDGE X-gates arise from reduced electric field screening from neighboring P-gates that do not overlap, as well as reduced dielectric material under X-gates with the uniplanar design. 

\newpage
\begin{figure}[h]
\includegraphics[width=1\linewidth]{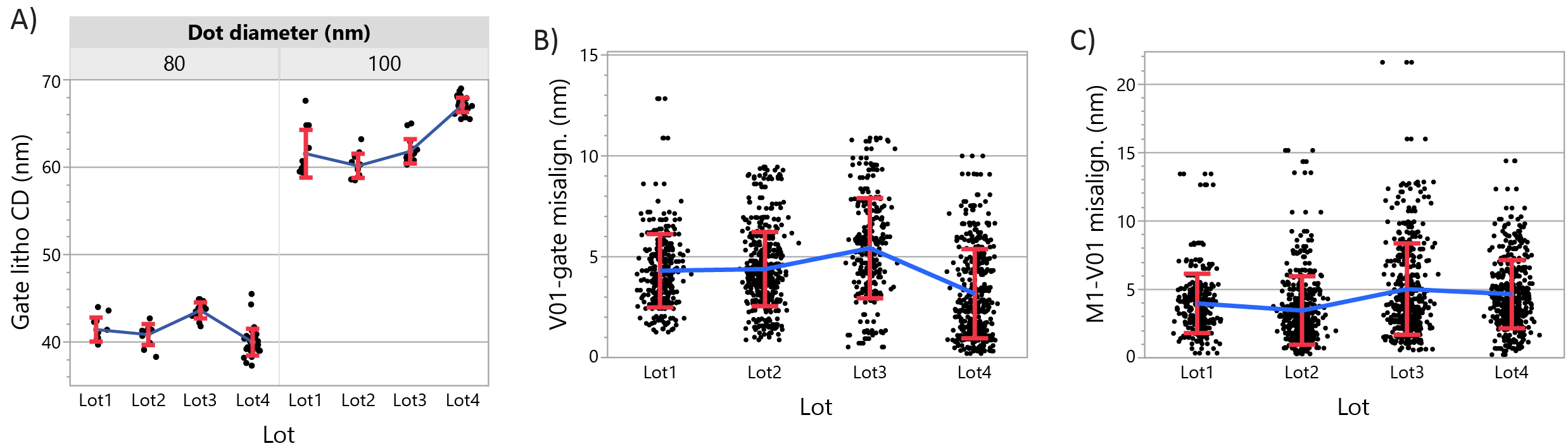}
\caption{(A) Plunger gate CD after e-beam lithography for two different dot diameters, across four lots.  (B) Magnitude of misalignment between V01 and gate levels across same four lots as in (A).  (C) Magnitude of misalignment between M1 and V01 levels across same four lots as in (A).  In all plots, blue lines connect lot means and red bars are one standard deviation from the mean.
}
\label{sfigcdsem}
\end{figure}

In Fig.~\ref{sfigcdsem}, we show the wafer-level reproducibility of e-beam lithography across lots for Si quantum dot fabrication, as measured by an automated critical dimension SEM (CDSEM) tool (Hitachi S9380).  As stated in the main text, each lot in this work consists of four \SI{75}{\mm} wafers patterned in one exposure sequence on a single \SI{200}{\mm} wafer holder.  Fig.~\ref{sfigcdsem}(a) shows the measured diameter in e-beam resist of plunger gates from two devices with nominally \SI{80}{\nm} and \SI{100}{\nm} designed dot diameter.  Although many devices are patterned on each wafer, we only measure the critical dimension (CD) from a small subset of devices and reticles.  It is evident in the chart that the lot-to-lot variation in mean lithography CD is $<7\%$ from the group mean.  

Figs.~\ref{sfigcdsem}(b-c) show the V01-gate and M1-V01 magnitudes of misalignment ($\sqrt{\Delta_x^2+\Delta_y^2}$) across the same four lots. Misalignment was measured at V01 and M1 e-beam lithography steps using box-in-box (BIB) structures wherein an interior rectangle is patterned in the prior etch step and an external rectangle is patterned in the current lithography step. The offset of the interior rectangle from the exact center of the exterior rectangle is the local misalignment. The CDSEM can rapidly extract $x$ and $y$ misalignments from \num{\sim100}s of BIBs from each wafer.  As seen in the figure, the mean misalignment magnitude for all lots is \SI{\lesssim5}{\nm} in both lithography steps.  The standard deviations are \SI{\le2.5}{\nm} for V01-gate BIBs and \SI{\le3.5}{\nm} for M1-V01 BIBs.  This level of misalignment across wafers and lots is well within the requirements necessary for SLEDGE devices, demonstrating that e-beam lithography is suitable for wafer-scale Si qubit fabrication.

\begin{figure}[h]
\includegraphics[width=0.8\linewidth]{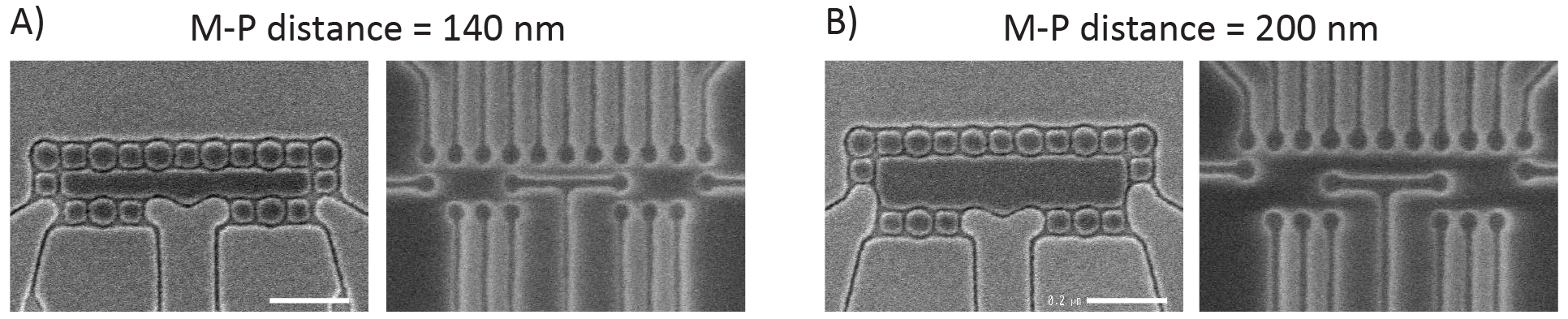}
\caption{Top-down SEM images at gate and M1 levels of devices with (A) \SI{140}{\nm} and (B) \SI{200}{\nm} M-P distance, with all other P- and X-gate dimensions fixed.  Note that the bath gates necessarily must extend vertically in (B) for longer M-P distance.  Scale bars correspond to \SI{200}{\nm}.
}
\label{sfigmpp}
\end{figure}

In the main text we discuss how the SLEDGE device platform enables straightforward tuning of P- and X-gate dimensions across devices. Another design parameter that can be easily modified within-reticle is the distance between P- and M-gate rows.  This dimension controls the sensitivity of M-gate charge sensors to electrons under corresponding P-gates (i.e. measurement signal-to-noise ratio).  If the M-P distance is too narrow, the Coulomb interaction from electrons under either M- or P-gates can adversely affect charge manipulation or sensing by the other, respectively.  In Fig.~\ref{sfigmpp}, we show top-down SEM images of two devices patterned within one reticle with M-P distance of \SI{140}{\nm} and \SI{200}{\nm} but all other dimensions fixed.  The M1 (and V01) pattern is easily adjusted in accordance with the gate-level design change. 

\newpage.
\begin{figure}[h]
\includegraphics[width=0.5\linewidth]{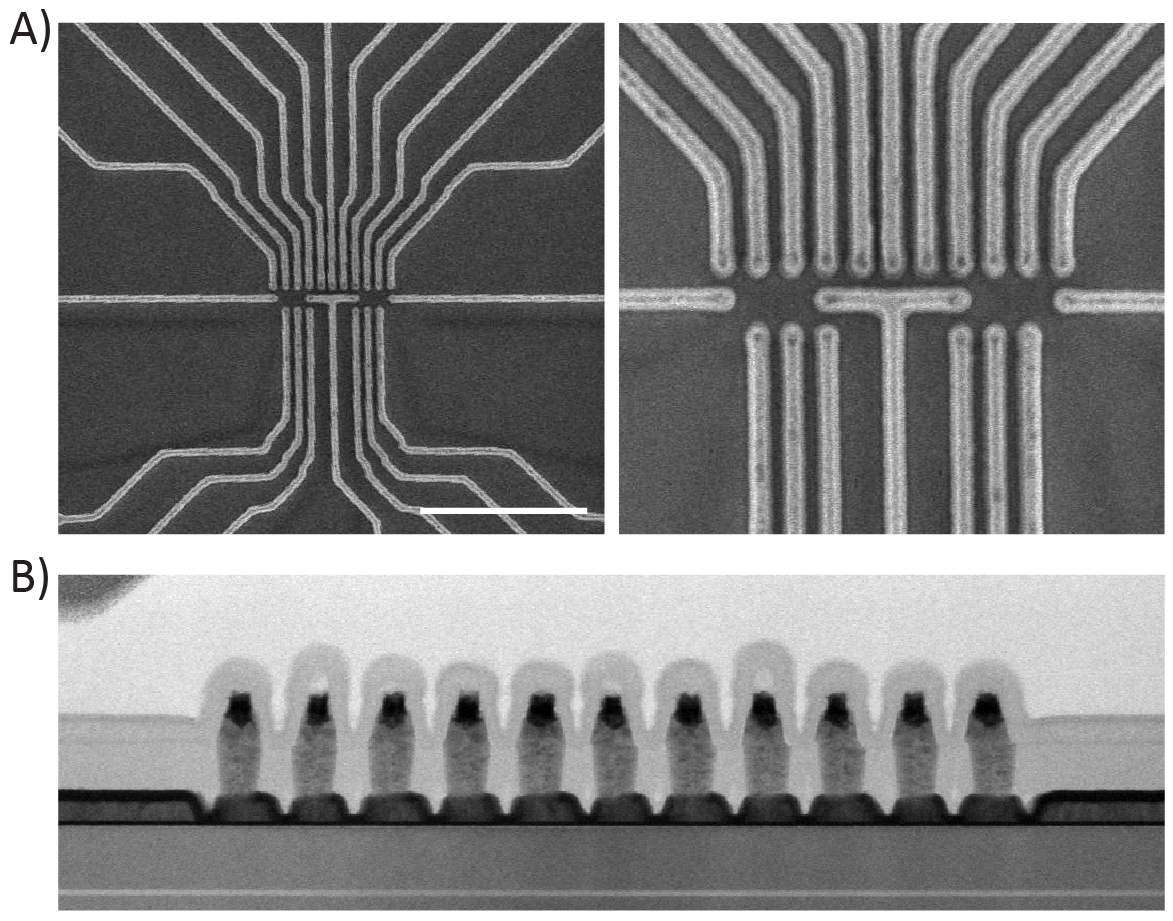}
\caption{(A) Top-down SEM images of M1 patterned by a subtractive etch process. Scale bar corresponds to \SI{1}{\um}. (B) Cross-section TEM image of a SLEDGE device with subtractive M1. The dark area on top of vias is W in M1, and the conformal gray layer over M1 is a protective \ce{SiN_x} encapsulation layer.
}
\label{sfigsub}
\end{figure}

The SLEDGE gate process can be used with a variety of BEOL integration schemes.  In Fig.~\ref{sfigsub} we show top-down SEM and cross-section TEM images from a subtractive etch process.  In this process, immediately after V01 etch, vias are filled by ALD TiN followed by blanket tungsten (W) sputter deposition.  Then the bilayer TiN/W stack is patterned by e-beam lithography and subtractively etched to form M1.  W serves as a low resistivity normal metal in the case that TiN superconductivity is broken by an applied magnetic field or by narrow linewidth. The subtractive process is advantageous over dual damascene because vias are filled immediately after etch, which avoids contamination in vias leading to poor via-gate connectivity.  Via-gate connectivity is discussed further below.

\begin{figure}[h]
\includegraphics[width=0.8\linewidth]{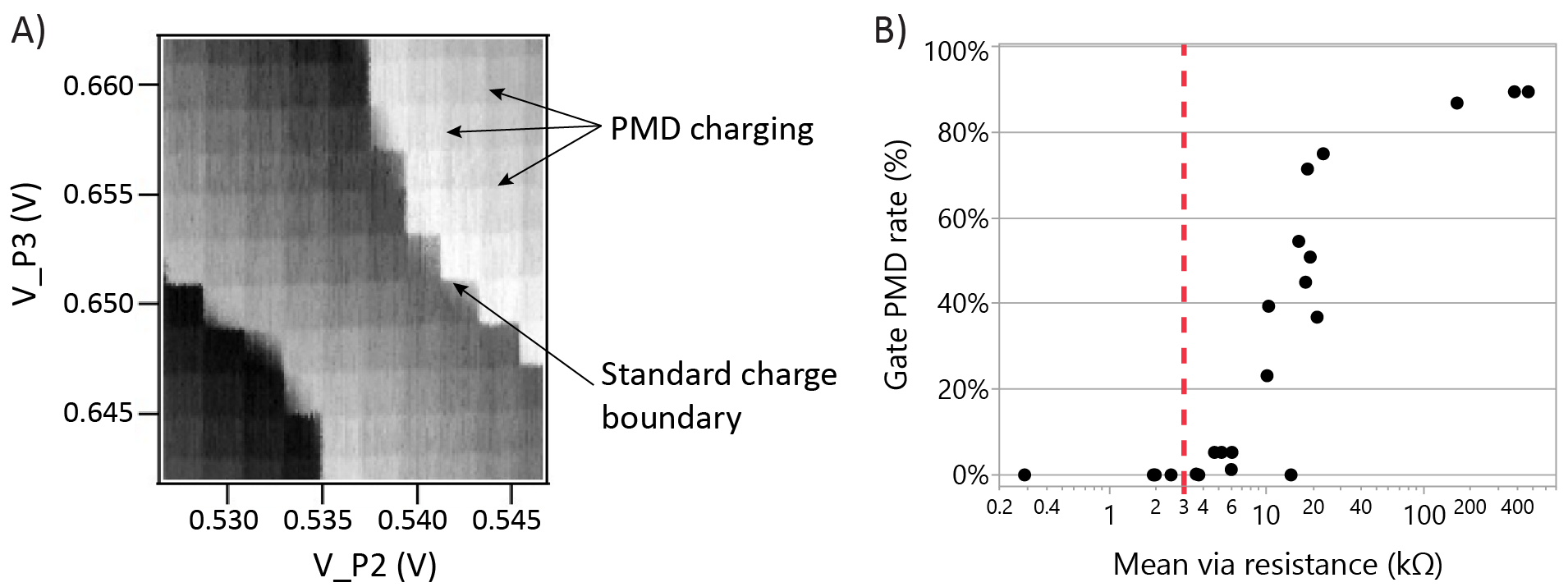}
\caption{(A) Charge stability diagram near boundary of (0,0) and (1,1) cells showing charging lines from parasitic metal dots.  (B) Gate-level PMD rate as a function of cryoprober mean via resistance.  Each data point represents one wafer.
}
\label{sfigpmd}
\end{figure}

In integrated circuits designed to operate at room temperature, a poor (non-Ohmic) connection between a via and a lower level metal feature may manifest as a high series resistance or open circuit.  In cryogenic quantum dot devices, an additional phenomenon may occur wherein a poorly connected gate acts as a floating quantum dot and electrons tunnel into the dot from vias across the non-Ohmic barrier.  This can be observed in charge stability diagrams as additional discrete, faint charging lines with slopes incommensurate with P-gate electron loading and with small addition voltages (Fig.~\ref{sfigpmd}(a)).  We denote this phenomenon ``parasitic metal dots" (PMDs), and it is considered a device failure if any gate exhibits PMDs.  We have developed a proxy metric for gate-level PMD rate based on via resistance measured by cryogenic on-wafer probing, which can provide more statistically meaningful wafer- and lot-level trends.  

In our fab-test workflow, after device fabrication but before \SI{1.6}{\K} electrostatic screening, whole wafers are electrically tested in a \SI{\sim40}{\K} automated on-wafer probing system (Cascade PAC200 with Celadon Systems probe card).  A representative subset of all devices on a wafer are pre-screened for isolation (\SI{\le100}{\pA} at \SI{1}{\V}) between all nearest-neighbor gates.  Other process control monitors such as Hall bars, capacitors, sheet resistance structures, transfer length method structures, via resistance structures, etc. are also measured.  This ``cryoprober" test provides a rapid (\SI{\sim1}{\day}) method for selecting plausibly good devices for further low temperature testing.  After die singulation, devices that were fully isolated in cryoprober measurements are packaged and wirebonded for \SI{1.6}{\K} screening.  

We can compare the mean via resistance measured per wafer in cryoprober against the gate-level rate of PMD observations in \SI{1.6}{\K} screening per wafer, as shown in Fig.~\ref{sfigpmd}(b).  The PMD rate is \SI{\sim0}{\percent} at low cryoprober via resistance and it nears \SI{\sim90}{\percent} at high via resistance.  Based on the number of nanoscale gates in a 6-dot device (20), the gate-level PMD rate should be as close to \SI{0}{\percent} as possible to not substantially impact device yield.  Based on the figure, we set a specification for SLEDGE cryoprober via resistance at \SI{3}{\kohm}.  In this manner, we can easily identify at cryoprober whether a wafer is expected to have a low PMD rate without having to use resources singulating die, packaging die, and measuring only finite number of devices at \SI{1.6}{\K}.  While in principle it should be possible to achieve low via resistance with any BEOL integration scheme, with our toolset we have had the greatest success with the subtractive process described above.  

\bibliography{qdotref}